\documentstyle[mncite,draft,epsf]{mn}

\newcommand{\lii}{Li\,{\footnotesize I}}

\newcommand{\cai}{Ca\,{\footnotesize I}}
\newcommand{\kms}{\,km\,s$^{-1}$}

\newcommand{\be}{\begin{equation}}
\newcommand{\ee}{\end{equation}}
\newcommand{\bd}{\begin{displaymath}}
\newcommand{\ed}{\end{displaymath}}

\title[Li in NGC 2547]{Cool stars in NGC 2547 and pre main sequence lithium depletion}  
  
\author[R. D. Jeffries et al.]{R. D. Jeffries$^{1}$\thanks{E-mail:
rdj@astro.keele.ac.uk},  
J. M. Oliveira$^{1}$, D. Barrado y Navascu\'{e}s$^{2}$, J. R. Stauffer$^{3}$ \\  
$^{1}$Astrophysics Group, Keele University, Keele, Staffordshire,  
ST5 5BG, UK\\  
$^{2}$ Laboratorio de Astrof\'{\i}sica Espacial y Fis\'{\i}ca
Fundamental, Apdo. 50727, 28080 Madrid, Spain\\
$^{3}$ SIRTF Science Center, California Institute of Technology, MS
314-6, Pasadena, CA 91125, USA}
\date{Received 31 Dec 2001}  
\pagerange{\pageref{firstpage}--\pageref{lastpage}}  
\def\LaTeX{L\kern-.36em\raise.3ex\hbox{a}\kern-.15em  
    T\kern-.1667em\lower.7ex\hbox{E}\kern-.125emX}

\begin{document}  
  
\label{firstpage}  
  
\maketitle  
  
\begin{abstract}  
We present the results of a spectroscopic survey of X-ray selected,
low-mass candidate members of the young open cluster NGC 2547. Using a
combination of photometry, spectroscopic indices and radial velocities
we refine our candidate list and then use our spectroscopy to study the
progression of lithium depletion in low-mass pre main sequence stars.
We derive lithium abundances or upper limits for all our candidate
members, which have effective temperatures in the range $5000>T_{\rm
eff}>3200$\,K, and compare these with predictions for lithium
burning and depletion provided by a number of models and also with the
lithium depletion seen in younger and older stars. We find that some
models {\it can} reproduce the lithium abundance pattern of NGC 2547 if
the cluster has an age of $\simeq 20-35$\,Myr, which is also indicated by
fits to low-mass isochrones in the Hertzsprung-Russell diagram. 
But the lack of significant further lithium depletion
between NGC 2547 and older clusters argues for  an age of at least
50\,Myr, more in keeping with the lack of lithium observed in even
fainter NGC 2547 candidates. We show that reconciliation of these age estimates
may require additions to the physics incorporated in current
generations of pre main sequence models.
\end{abstract}  
  
  
\section{Introduction}  

Tracing the evolution of lithium abundances in the cool stars of
Population I open clusters is an excellent way of exploring internal
mixing in stars with partially or fully convective envelopes. In the
last 15 years, many Li abundance measurements have been made in dozens
of clusters (see Jeffries 2000 and references therein). An empirical
picture has emerged from this observational work that is at odds
with the parallel efforts of theory.

``Standard'' models incorporate convection as the sole means of mixing
Li-depleted material from hot interior layers to the stellar surface.
Photospheric Li depletion occurs swiftly once the base of the
convection zone (or if fully convective, the core of the star) becomes
hot enough to burn Li in p,$\alpha$ reactions (at $\sim
2.5\times10^{6}$\,K). For G and K stars with $T_{\rm eff}>4500$\,K, Li
depletion should have largely ended by the time these stars reach the
ZAMS, because their convection zone bases have pushed outwards to
cooler temperatures (Pinsonneault 1997). Yet the Sun's Li abundance is
nearly two orders of magnitude lower than ZAMS stars of similar mass,
and observations of open clusters with different ages clearly show (see
Thorburn et al. 1993; Soderblom et al. 1993; Jones, Fischer \&
Soderblom 1999; Ford et al. 2001) that Li depletion continues during
main sequence evolution, on timescales of 100 Myr to a few Gyr.
Furthermore, the expected strong metallicity dependence of PMS Li
depletion is not present. Clusters with the same age, but quite
different metallicity have similar Li depletion patterns among their F,
G and K stars when they reach the ZAMS (Jeffries \& James 1999; Barrado
y Navascu\'{e}s, Deliyannis \& Stauffer 2001; Jeffries et al. 2002).

\begin{figure*}
\vspace*{9cm}
\includegraphics{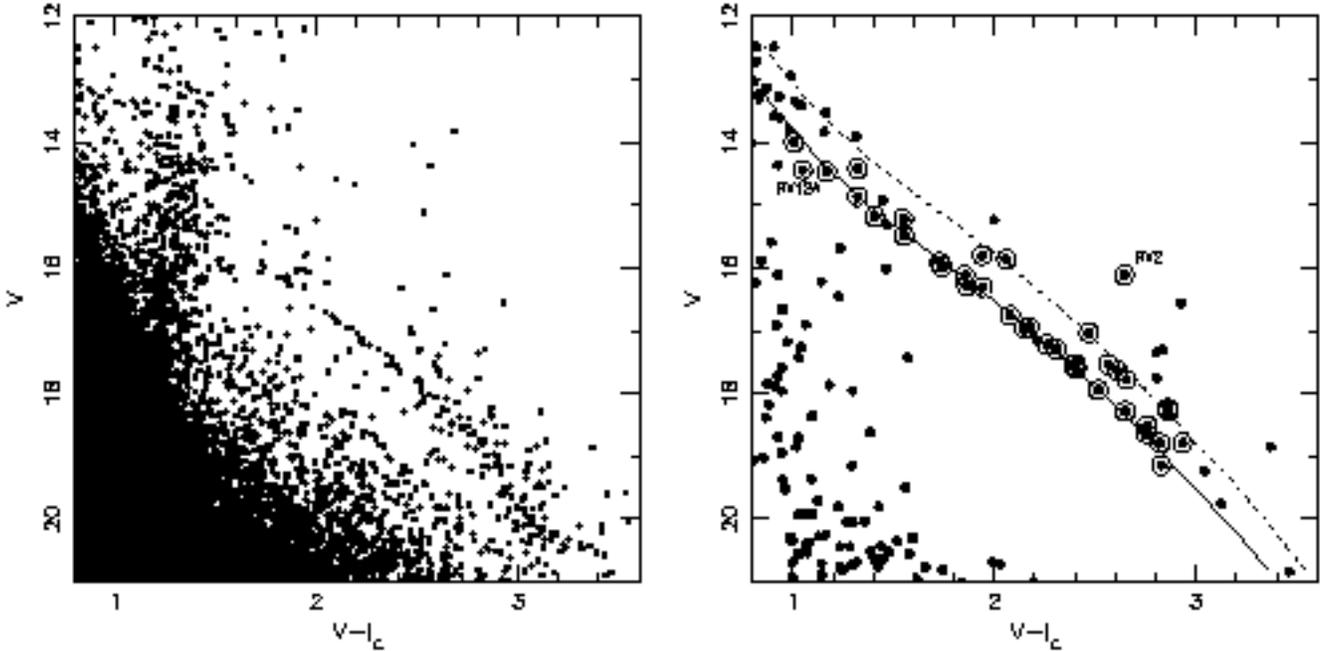}
\caption{(a) The $V$,$V-I_{\rm c}$ diagram for objects within the field
of view of the {\em ROSAT} X-ray observation of NGC 2547. (b) A similar
diagram showing only those objects correlated with an X-ray source. We
have circled those objects for which spectroscopic data have been
obtained in this paper. The solid line is a 25\,Myr isochrone derived
from the models of Baraffe et al. (1998, 2002 -- see
Sect.~\ref{discuss}) at an intrinsic distance modulus of 8.15 and the
dashed line is where binary systems with equal mass (and hence
brightness) components would lie.}
\label{vvin2547}
\end{figure*}

So far most work has concentrated on the warmer F, G and K stars, primarily
because they are brighter and have higher Li abundances. Li is
undetectable in the Pleiades (age 120 Myr) below about 4200\,K (Jones
et al. 1996), but is clearly seen in low mass PMS stars of similar
temperatures in star forming regions at ages of a few Myr
(e.g. Mart\'{\i}n et al. 1994; Zapatero Osorio et al. 2002). Rapid PMS
Li-destruction in these cool stars is qualitatively expected from
``standard'' evolutionary models but the detailed behaviour is quite
dependent on a number of uncertain features in stellar evolution models
-- particularly the treatment of convection and convective overshoot
(Chaboyer, Demarque \& Pinsonneault 1995). A recent paper by Randich et
al. (2001) has presented Li measurements in cool K and M
dwarfs of the young (age 30-50\,Myr) IC 2391 and IC 2602 clusters. They
find that these stars are depleted from their initial Li abundance and
are on average more Li rich than stars in the Pleiades. However, some K
stars have Li abundances that are nearly as low as their most Li-depleted
Pleiades counterparts. Randich et al. deduced that unless Li depletion was halted,
these stars would ultimately become more Li-depleted than Pleiades
stars of the same effective temperature ($T_{\rm eff}$), by the time
they reached the ZAMS. This conclusion is surprising because, as
mentioned above, no significant differences in the level of Li
depletion have been seen among the G and K stars of several open
clusters with age $\sim 100$\,Myr.

In this paper we present the results of an investigation into Li
depletion among the cooler K and M stars of NGC 2547 (=C0809-491), a
young cluster at a distance of $\simeq 400$\,pc and with a small
reddening, $E(B-V)=0.06$ (Clari\'a 1982).
Jeffries \& Tolley (1998) found that NGC 2547 was a very young cluster,
using an X-ray survey that discovered many coronally active low-mass
cluster members. Fitting low-mass isochrones, Naylor et al. (2002) have
determined an age of 20-35\,Myr for NGC 2547. We have used fibre
spectroscopy to confirm cluster membership and 
study the \lii\ 6708\AA\ resonance line in a sample of
X-ray selected low-mass stars in NGC 2547. Section~\ref{target}
describes our target selection and the
spectroscopy. Section~\ref{results} presents our results in terms of
spectral classification, radial velocities and the strength of the H$\alpha$ and \lii\
6708\AA\ lines. In Sect.~\ref{contaminate} we address the issues of
completeness and contamination in our sample. In Sect.~5
we estimate Li abundances for the cluster members and compare them with
theoretical models and similar data for the Pleiades, IC 2391 and
IC~2602. Our results are discussed in Sect.~6.

\section{Target Selection and Spectroscopy}

\label{target}

Targets for fibre spectroscopy were taken from Table~2 of Jeffries
\& Tolley (1998) and we refer to them throughout this paper by the
identifier given there. These stars were identified as responsible for
the 0.1-2\,keV X-ray emission seen by the {\em ROSAT} satellite and
almost all were considered likely members of NGC 2547 on the basis of
their positions in $BV$ (Johnson), $I_{\rm c}$ (Cousins)
colour-magnitude and colour-colour
diagrams\footnote{In this paper we will use magnitudes and colours for
our targets taken from Naylor et al. (2002). These can differ at the level
of a few hundredths of a magnitude from those in Jeffries \& Tolley
(1998), but should be more precise.}.  The targets range in brightness
from $13.99\leq V \leq19.15$, with a colour range of $1.01\leq V-I_{\rm
c} \leq 2.93$ and a mass range of approximately $0.3<M<0.9M_{\odot}$
(using the models of Baraffe et al. 1998, 2002). 
Figure~\ref{vvin2547}b shows the $V$ versus
$V-I_{\rm c}$ diagram for our targets.

As NGC 2547 is relatively compact, we were able to use the 2dF
spectrograph at the 3.9-m Anglo-Australian Telescope (see Lewis et
al. 2002) to obtain spectra of 37 objects simultaneously.  Very
accurate astrometry for our targets was obtained firstly from our own
CCD images (Jeffries \& Tolley 1998; Naylor et al. 2002). Positions were
then improved by comparison with scans of the appropriate UK Schmidt
plate from the SuperCOSMOS sky survey (see Hambly et al. 2001). Fifteen
fibres were placed on adjacent blank-sky areas that had no stars visible on
the Schmidt plates. The observations were performed in service mode on
24 January 1999. The 1200V grating was used at a central wavelength of
6800\AA. Unfortunately this was not quite the optimal configuration for our
purposes -- the 1200R grating has a somewhat higher efficiency at these wavelengths.
A wavelength range of approximately 1080\AA\ and a 2.5-pixel resolution of about 2.7\AA\
were achieved. Four separate target exposures were obtained, each of length 1800\,s,
together with tungsten and arc lamp exposures and two
300\,s exposures of offset ``blank'' sky.

Data reduction was achieved using version 2.0 of the {\sc 2dfdr}
package (Bailey, Glazebrook \& Bridges 2002), which included bias
subtraction, removal of scattered light and optimal extraction of the
spectra.  Relative
fibre transmissions were determined from offset sky exposures. A mean
sky spectrum was calculated from the fifteen sky fibres and a scaled
version of this was subtracted from each target spectrum. By looking at
the results of sky-subtraction for the sky fibres themselves, we
estimate that the relative fibre transmissions were accurate to about five
per cent.

\section{Results}

\label{results}

\subsection{Spectral classification}

\label{specclass}

For each target we calculated two narrow-band spectral indices, both of which are temperature
sensitive and the second of which is somewhat sensitive to surface
gravity. These indices are defined by Brice\~no et al. (1998, following
Kirkpatrick, Henry \& McCarthy 1991; Allen 1995) as
\bd
{\rm TiO(7140\AA)} = \frac{C(7020-7050{\rm \AA})}{T(7125-7155{\rm \AA})}\, ,
\ed
\bd
{\rm CaH(6975\AA)} = \frac{C(7020-7050{\rm \AA})}{T(6960-6990{\rm \AA})}\, ,
\ed
where $C$  and $T$ indicate pseudo-continuum and absorption features,
integrated over the indicated wavelength ranges.

The indices were calibrated using a set of older late K and M dwarfs in
the Praesepe open cluster (age\,$\simeq800$\,Myr) observed by Allen \&
Strom (1995), a set of M dwarf field stars observed at $\sim 2-3$\AA\
resolution on the Keck II and 4-m Blanco telescopes (see Barrado y
Navascu\'{e}s, Stauffer \& Patten 1999) and K and M dwarf field stars
taken from the spectral atlas of Montes et al. (1997). To ensure
comparability, all spectra were blurred to match the 6\AA\ resolution
of Allen \& Strom's data before calculating the indices. Photoelectric
$V-I_{\rm c}$ photometry is available for most of the field stars from
Stauffer \& Hartmann (1986), Bessell (1990) or Leggett (1992).  Colours
on the Kron system were converted to colours on the Cousins system
using formulae from Bessell \& Weis (1987).

\begin{figure}
\vspace*{10cm}
\includegraphics{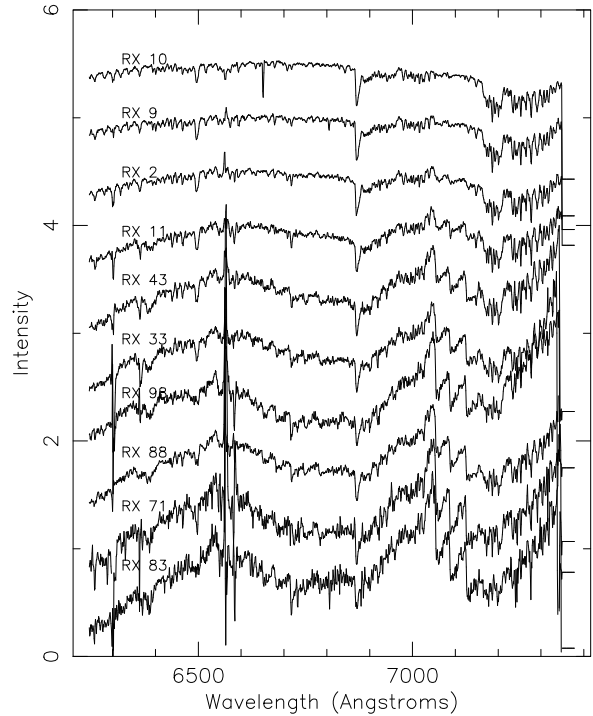}
\caption{A selection of our target spectra exhibited in approximate
spectral type order.}
\label{speccy1}
\end{figure}
\begin{figure}
\vspace*{7.7cm}
\includegraphics{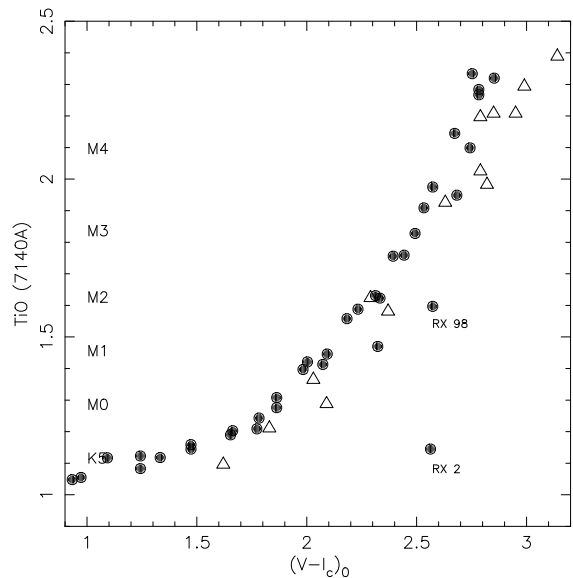}
\caption{The TiO(7140\AA) narrow band index as a function of intrinsic
colour for our spectroscopic targets (spots), compared with field
stars. Spectral types are allocated according to the strength of the
TiO index as shown along the y-axis.}
\label{vitplot}
\end{figure}
\begin{figure}
\vspace*{7.7cm}
\includegraphics{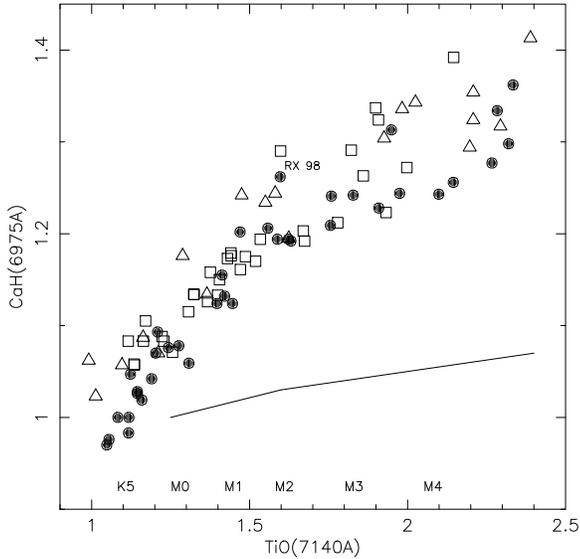}
\caption{The CaH(6975\AA) versus TiO(7140\AA) narrow band indices for
our spectroscopic targets (spots), compared with similar data for a
sample of stars in Praesepe (squares -- from Allen \& Strom 1995) and
field stars (triangles). The solid line indicates the locus of giants
in this plot found by Allen \& Strom.}
\label{tcplot}
\end{figure}

We use the TiO(7140\AA) index as the primary spectral type indicator.
The spectral types listed in Table~\ref{tab1} are estimated to be good
to the nearest subclass. In Fig.~\ref{vitplot}, we show the
TiO(7140\AA) index as a function of intrinsic $V-I_{\rm c}$ colour for
NGC 2547 and the field stars. An $E(V-I_{\rm c})$ of 0.077 was assumed
for NGC 2547 (see Jeffries \& Tolley 1998).  The plot shows reasonable
agreement, with perhaps a hint that the $V-I_{\rm c}$ colours in NGC
2547 are too blue by $\sim0.05-0.1$ mag. Systematic errors in the
photometric calibration at this level are not surprising given the
difficulty in finding very red calibration stars.  The stars RX\,2 and
RX\,98 stand out as unusual objects. Their colours are too red for
their spectral types. Figure~\ref{speccy1} shows some of our spectra in
order of spectral type, including those of RX\,2 and RX\,98.

Figure~\ref{tcplot} shows the gravity sensitive CaH(6975\AA) index
plotted aginst the TiO(7140\AA) index. This time we have included the
Praesepe stars as well, because this plot does not require accurate
$V-I_{\rm c}$ photometry. Also shown is a locus (from Allen \& Strom
1995), indicating where low gravity giants lie.  All our targets have
dwarf-like gravities, although there is a hint that the gravities are
lower on average than in Praesepe and the field stars. This is
precisely what we would expect from a population of late K and M dwarfs
in a cluster of age $\sim30$\,Myr. The evolutionary models of Chabrier \& Baraffe
(1997) indicate that stars of 0.3 to
0.9$M_{\odot}$ and age $\simeq 30$\,Myr, have surface gravities that are between
0.46 and 0.11 dex lower than at 800\,Myr, but which are still several
dex higher than M giant gravities.

\subsection{Relative radial velocities}

Genuine members of NGC 2547 should share a common radial velocity 
(subject to the intrinsic cluster dispersion of $<0.9$\kms\ -- Jeffries, Totten
\& James 2000), unless
they are in close binary systems. Unfortunately, we could not obtain
spectra of radial velocity (RV) standards during our service
observations and so cannot put our
observations onto an absolute heliocentric velocity scale. Nevertheless we can
find {\em relative} RVs for our targets by cross-correlation. 

The targets were split into three groups according to their spectral
types -- K, M0-M2 and M3-M4. For the K-stars we have three targets,
RX\,3, 10 and 12A, for which heliocentric RVs were measured by Jeffries
et al. (2000). RX\,12A was measured twice on separate nights and found
to have a constant RV that is {\em inconsistent} with cluster membership (it is a
photometric non-member as well -- see Sect.~\ref{member}).  We choose
this star to act as our primary standard and cross-correlate it against
the other K-stars, using the wavelength range $\lambda\lambda
6310-6530$\AA, which contains no significant telluric absorption or sky emission
lines. Apart from RX\,76A, we find that the K-stars have an RV that is
$14.9\pm2.0$\kms\ smaller than RX\,12A, with an rms scatter of 
4.9\kms. A large fraction of this scatter is likely due to the precision of
our wavelength calibration. A typical fibre exhibited a 10\kms\ rms
between the measured and calculated positions of the lines. Given that
about 20 lines were used in the calibration solution, this suggests
zeropoint uncertainties of just over 2\kms\ in a single measurement and about 3\kms\
in a cross-correlation lag, assuming that the target and template match perfectly.
The mean RV is in excellent agreement with Jeffries et al. (2000),
where the heliocentric RV of RX\,12A was found to be
$(28.9\pm0.5)$\kms\ compared with the cluster mean of
$(12.8\pm0.2)$\kms. Thus most of the K stars can be classified as
cluster members based on their RV (including RX\,2, which is a clear
photometric non-member).  For RX\,76A we find an RV which is 0.8\kms\
greater than that of RX\,12A.

As we move towards cooler stars, molecular features becomes more
prominent. For these objects we cross-correlated the wavelength range
$\lambda\lambda6950-7150$\AA, containing the strong CaH and TiO
features discussed in Sect.~\ref{specclass}. For the M0-M2 stars we
used RX\,81 as the template and for the M3-M4 stars we chose RX\,90. In
both groups we find strong clustering of relative RVs around zero, the
only significant exception being RX\,60.  The mean RV of the M0-M2
stars relative to RX\,81 is $-1.2$\kms, with a standard deviation of
4.4\kms\ (from 15 stars), and the mean RV of the M3-M4 stars relative
to RX\,90 is $-3.4$\kms\ with a standard deviation of 4.5\kms\ (from 10
stars). RX\,60 is clearly discrepant with an RV (relative to RX\,90) of
$-15.7$\kms.

Whilst RV is capable of identifying non-members of the cluster, it is
not conclusive at this level of precision.  As our
objects are X-ray selected, we expect any contaminating field objects
to be X-ray active, have young disk kinematics and hence RVs that are
within 20-30\kms\ of NGC 2547. Nevertheless it is encouraging that all
but three of our targets have RVs that are within 10\kms\ of the mean
and by selecting on RV we should cut down the level of possible 
contamination by a factor of a few.

\subsection{H$\alpha$ equivalent widths}

\label{halphaew}

The contribution from the ``sky'' to H$\alpha$ in the target fibres is
substantial for our targets and exceeds the target signal by factors
of ten in some cases. Because of this, uncertainties in the
relative fibre throughputs are important, however the dominant source
of error is the strong, spatially varying H$\alpha$ emission from an
H\,{\sc ii} region that appears to coincide with NGC 2547 and is
visible even on the Digitized Sky Survey images. With only 15 sky
fibres, we have not sampled this emission finely enough to provide
anything but a crude sky subtraction for the stellar targets.
The H$\alpha$ equivalent widths (EWs) were simply determined
by direct integration above a pseudo-continuum. The uncertainties have
been estimated by assuming they are similar in size to the residuals
seen in spectra measured from the sky fibres at H$\alpha$ {\em after} the mean sky
has been subtracted. This residual flux is combined with the observed continuum
flux of the stellar targets to form an EW uncertainty.

The H$\alpha$ EWs are given in Table~\ref{tab1}. Even given the
substantial uncertainties, many of our targets show significant H$\alpha$
in emission, indicative of chromospheric activity. None of the targets 
exhibit an H$\alpha$ feature that is inconsistent with
young, magnetically active members of NGC 2547. 

\subsection{Lithium 6708\AA\ equivalent widths}

\begin{figure}
\vspace*{7cm}
\includegraphics{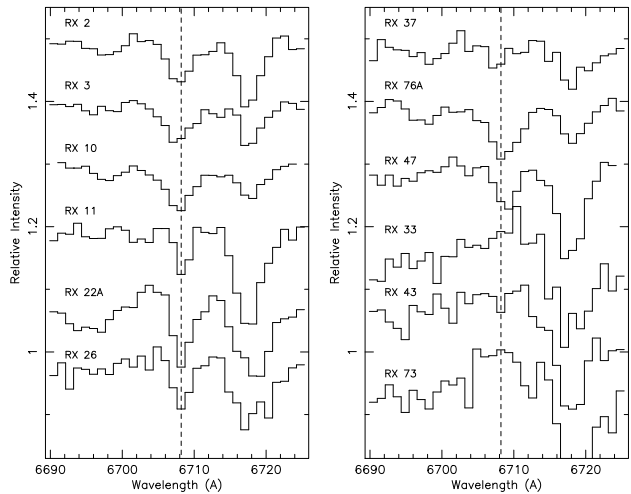}
\caption{Examples of our spectra around the \lii\ 6708\AA\ feature. We
also show three examples (RX\,33, RX\,43, RX\,73) where only upper
limits to the \lii\ pEW could be found. The spectra have been
normalised and displaced in steps of 0.1 for clarity.}
\label{lispecplot}
\end{figure}

\label{liew}

There are no sky lines present at the expected position of the 6708\AA\
\lii\ absorption feature. The sky continuum level at this wavelength is
smaller than the continuum levels of all our targets and so the small
uncertainty in the relative fibre throughput calibration is of little
concern. Some examples of our spectra in the \lii\ 6708\AA\ region are
shown in Fig.~\ref{lispecplot}.

We define two pseudo-continuum regions either side of the line and then
integrate the \lii\ line over a fixed wavelength range to
determine a pseudo-EW (pEW). The (internal) uncertainties are dominated
by the signal-to-noise ratio per pixel (SNR) of our spectra. These are
empirically (and conservatively) determined from the rms to straight
line fits to our continuum regions. The pEW error is then estimated
from the equation, $\delta ({\rm pEW}) \simeq (\sqrt{r\, p})/{\rm SNR}$, where
$r$ and $p$ are the width over which the line is integrated and the
width of a pixel respectively. Where no significant \lii\ line is seen,
we estimate a 2-sigma upper limit. These pEWs (and upper limits) are
listed in Table~\ref{tab1}.

\subsection{Final membership status}

\label{member}

The membership status of each target is listed in col.~8 of
Table~\ref{tab1}. We are limited to using the photometry, RVs
and spectral-types to classify them. Contaminating field
stars in an X-ray selected sample are likely to be chromospherically
active and possibly Li-rich in the case of K-stars (e.g. Jeffries 1995), 
so we should not use the presence of H$\alpha$
emission or Li as membership criteria.  Initially we check the position
of each star in the $V,V-I_{\rm c}$ colour-magnitude diagram (CMD -- see
Fig.~\ref{vvin2547}b) against the expectations of an isochrone
generated from the Baraffe et al. (1998) models (see
Sect.~\ref{discuss}), that appears to fit the bulk of the targets very
well. If an object lies between 0.3 and 0.9 mag above this isochrone we
class it as a probable binary system. The likely mass ratio of such
systems would be $0.6<q<1$. There are two objects (RX\,2 and RX\,12A)
which we classify as non-members on the basis of their photometry.

Next we compare the spectroscopic classification with the $V-I_{\rm c}$
colour. From this we judge that RX\,2\footnote{The spectral type and
colour of RX\,2 are difficult to reconcile. This, together with a
strong \lii\ 6708\AA\ feature and RV consistent with cluster
membership, suggest this object may be a cluster member with a
weak cosmic ray affecting its I-band photometry (i.e. it could be a
K-type cluster member with an anomalous $V-I_{\rm c}$ value).}
and RX\,98 are probably not cluster members. Finally we note that
RX\,12A, RX\,60 and RX\,76A were found to have RVs discrepant from the
cluster mean. We have already discarded RX\,12A as a member on
photometric grounds, but the other two objects may be cluster members
in binary systems and RX\,76A does show photometric evidence of
binarity. We treat RX\,60 and RX\,76A as {\it possible} cluster members
in what follows.

\begin{table*}
\caption{The photometry and spectroscopic properties of our
targets. Columns 1-5 are self-explanatory, col.6 lists the
pseudo-equivalent width of the \lii\ feature (see Sect.~\ref{liew}), col.7 gives the
signal-to-noise ratio of the spectra around the \lii\ line and col.8
records the membership status discussed in Sect.~\ref{member}. S
indicates a single cluster member (or at least no evidence of
binarity), B indicates an object lying more than 0.3 mag. in $V$ above
the cluster single star isochrone, NM indicates that we consider the
target a cluster non-member, either because of its photometry (p), its
spectral type compared with its colour (s) or because of its discrepant
RV (r -- although these objects are treated as possible short period binary
cluster members). Columns 9 and 10 list effective temperatures and
derived LTE Li abundances for the cluster members (Sect.~5.1).}
\begin{tabular}{rcccrrclcc}
\hline
Name & $V$ &$V-I_{\rm c}$& SpT & H$\alpha$ EW&\lii\ pEW & SNR & Status & $T_{\rm eff}$ & A(Li) \\ 
RX   &     &             &     & (\AA)       & (m\AA)   &   &  &  (K) &   \\
 2   &16.113 &2.644 &K7 & $  0.7\pm0.4$ &$  220\pm24$ & 100 &NM(p,s)
& 3445 & -           \\
 3   &14.419 &1.324 &K7 & $  0.3\pm0.4$ &$  216\pm24$ & 100 &B
& 4343	& $1.58^{+0.21}_{-0.22}$\\
 9   &14.863 &1.321 &K5 & $  0.3\pm0.3$ &$  140\pm22$ & 120 &S
& 4347	& $1.16^{+0.26}_{-0.31}$\\
10   &13.993 &1.006 &K3 & $ -0.1\pm0.2$ &$  214\pm17$ & 190 &S
& 4983	& $2.66^{+0.21}_{-0.21}$\\
11   &16.120 &1.854 &M0 & $  2.4\pm0.8$ &$  160\pm39$ &  60 &S
& 3804	& $0.50^{+0.57}_{-0.41}$\\
12A  &14.441 &1.049 &K3 & $ -0.4\pm0.3$ &$  147\pm18$ & 130 &NM(p,r)
& 4880	& -          \\
14   &16.951 &2.173 &M1 & $  0.30\pm1.5$ &$ < 70     $ &  70 &S
& 3660	& $<0.00$     \\
17   &17.291 &2.305 &M2 & $  3.0\pm2.0$ &$ < 65     $ &  75 &S
& 3609	& $<0.00$     \\
21A  &17.543 &2.566 &M3 & $  2.9\pm2.2$ &$ <120     $ &  40 &B
& 3490	& $<0.00$     \\
22A  &15.863 &2.060 &M1 & $  2.3\pm0.70$ &$  363\pm39$ &  60 &B
& 3705	& $1.90^{+0.33}_{-0.34}$\\
23   &19.147 &2.827 &M4 & $  6.7\pm10.1$ &$ <490     $ &  10 &S
& 3322	& $<2.60$     \\
25A  &18.245 &2.859 &M4 & $  5.8\pm4.4$ &$ <330     $ &  15 &B
& 3297	& $<1.34$     \\
26   &16.301 &1.940 &M0 & $  3.7\pm0.9$ &$  241\pm78$ &  30 &S
& 3758	& $1.01^{+0.63}_{-0.61}$\\
33   &17.223 &2.262 &M2 & $  6.4\pm1.4$ &$ < 75     $ &  65 &S
& 3626	& $<0.00$     \\
37   &15.915 &1.728 &M0 & $  3.1\pm0.8$ &$  110\pm39$ &  60 &S
& 3886	& $0.01^{+0.19}_{-0.50}$\\
40   &17.546 &2.398 &M1 & $  3.6\pm2.2$ &$ <165     $ &  30 &S
& 3571	& $<0.26$     \\
43   &16.966 &2.149 &M1 & $  2.4\pm1.5$ &$ < 70     $ &  70 &S
& 3670	& $<0.00$     \\
44   &18.519 &2.755 &M4 & $  1.6\pm5.8$ &$ <195     $ &  25 &S
& 3373	& $<0.47$     \\
46A  &18.796 &2.820 &M4 & $  8.5\pm5.7$ &$ <330     $ &  15 &S
& 3327	& $<1.38$     \\
47   &16.279 &1.864 &M0 & $  1.6\pm0.8$ &$  197\pm36$ &  65 &S
& 3798	& $0.78^{+0.34}_{-0.29}$\\
50A  &15.484 &1.554 &K7 & $  0.8\pm0.5$ &$  168\pm29$ &  80 &S
& 4040	& $0.92^{+0.17}_{-0.17}$\\
60   &17.943 &2.515 &M3 & $  0.5\pm4.2$ &$ <140     $ &  35 &S,NM(r)
& 3515	&  $<0.03$      \\
61B  &17.608 &2.406 &M2 & $  1.6\pm2.4$ &$ <165     $ &  30 &S
& 3567	& $<0.26$     \\
65A  &16.762 &2.078 &M1 & $  1.5\pm1.30$ &$  100\pm58$ &  40 &S
& 3698	& $<0.35$\\
71   &18.644 &2.753 &M4 & $ 18.5\pm5.5$ &$ <330     $ &  15 &S
& 3375	& $<1.44$     \\
73   &17.553 &2.392 &M2 & $  1.8\pm2.1$ &$ <110     $ &  45 &S
& 3573	& $<0.00$     \\
76A  &15.224 &1.549 &K7 & $  2.3\pm0.4$ &$  343\pm25$ &  95 &B,NM(r)
& 4046	& $2.12^{+0.21}_{-0.19}$  \\
80A  &15.180 &1.407 &K5 & $  0.1\pm0.4$ &$  137\pm24$ & 100 &S
& 4219	& $0.89^{+0.14}_{-0.23}$\\
81   &15.981 &1.739 &M0 & $  1.4\pm0.70$ &$  142\pm39$ &  60 &S
& 3878	& $0.57^{+0.73}_{-0.33}$\\
82   &14.467 &1.168 &K5 & $  0.1\pm0.20$ &$  155\pm24$ & 100 &S
& 4622	& $1.70^{+0.17}_{-0.20}$\\
83   &18.793 &2.933 &M4 & $  3.3\pm6.2$ &$ <245     $ &  20 &S
& 3238	& $<0.69$     \\
84B  &18.285 &2.860 &M4 & $  3.9\pm4.0$ &$ <165     $ &  30 &B
& 3296	& $<0.14$     \\
88   &17.037 &2.470 &M3 & $  2.9\pm1.3$ &$ <110     $ &  45 &B
& 3537	& $<0.00$     \\
90   &18.278 &2.648 &M4 & $  4.6\pm 4.0$ &$ <165     $ &  30 &S
& 3443	& $<0.26$     \\
93   &17.646 &2.612 &M4 & $  6.0\pm2.1$ &$ <100     $ &  50 &B
& 3464	& $<0.00$     \\
95   &15.803 &1.939 &M0 & $  1.2\pm1.3$ &$ < 80     $ &  60 &B
& 3759	& $<0.00$     \\
98   &17.779 &2.654 &M2 & $  2.9\pm2.7$ &$ <165     $ &  30 &NM(s)
& 3439	& - \\
\hline
\end{tabular}
\label{tab1}
\end{table*}

\section{Completeness and Contamination}

\label{contaminate}

Our sample is certainly not complete however this is unlikely to be a 
problem in interpreting Li depletion patterns unless the sample is somehow 
unrepresentative. A possible problem is that,
through X-ray selection, our sample is biased towards more rapidly
rotating and magnetically active stars, which may not suffer the same
Li-depletion as slowly rotating stars.  The sample may also
be contaminated with field objects.  Jeffries \& Tolley (1998)
demonstrate that only one of the optical counterparts to X-ray sources
coincident with the NGC 2547 PMS locus might be a chance
correlation. Therefore contaminating field stars would
also be X-ray active.

Figure~\ref{vvin2547}a shows that subset of the CMD
for NGC 2547 (from Naylor et al. 2002) that coincides with the 17
arcmin radius {\em ROSAT} observation of Jeffries \& Tolley (1998 --
see their Fig.~1). In Figure~\ref{vvin2547}b we show those objects
close to an X-ray source and circle targets
for which we have spectra in this paper.  We have not
observed every X-ray counterpart that has photometry consistent with
NGC 2547 membership, but we have observed nearly all of them with
$1.0<V-I_{\rm c}<3.0$.  The discussion of completeness and
contamination among the spectroscopic targets is most conveniently
split into three colour ranges.

$1.0<V-I_{\rm c}<1.5$: In this colour range, contamination among a
photometrically selected sample is expected from background
giants.  Fortunately, these contaminants are not strong X-ray sources,
so an X-ray selected sample should not contain giants.  Contamination
by young, active field dwarfs is a possibility, but the low density of
X-ray sources both above and below the cluster sequence suggests there
are few such contaminants in our entire sample.  Furthermore, Jeffries
\& Tolley (1998 - their Sect.~4.2) argue that X-ray active, young field
dwarfs would {\em tend} to lie well below the cluster PMS locus at
these colours.  The X-ray selected sample of cluster members in this
colour range is small but also likely to be complete over the X-ray
survey area. This is because the cluster candidates coincident with
X-ray sources at these colours have X-ray luminosities far higher than
the threshold for detection (see Jeffries \& Tolley 1998; Jeffries et
al. 2000).  So our X-ray selected sample is complete and other (X-ray
quiet) stars in the same region of the CMD are likely to be contaminating 
background giants.

$1.5<V-I_{\rm}<2.5$: The low density of stars in the CMDs,
both above and below the cluster sequence suggest that a
photometrically selected sample of cluster candidates will suffer
little contamination in this colour range (see Naylor et al. 2002). If
we further restrict the sample to X-ray selected objects, then any
small amount of contamination, mainly from field dwarfs, would be
reduced by another factor of $\sim 10$, since only field stars with
ages\,$\la 1$ Gyr would be so X-ray active. However, an X-ray selected
sample is unlikely to be complete in this colour range, with only the
most X-ray active NGC 2547 members being detected by {\em ROSAT}. If we
assume that the photometric members in Fig.\ref{vvin2547}a suffered no
contamination, then the X-ray selected sample is about
50 per cent complete, becoming less so towards redder colours.

$V-I_{\rm c}>2.5$: The PMS locus of NGC 2547 appears to
broaden here because of increasing photometric errors
and variability for late dM stars. The cluster sequence
is less clear and it becomes more difficult to separate cluster and
field stars solely from photometry. The X-ray sources in this
region of the CMD have $L_{\rm x}/L_{\rm bol}\sim10^{-3}$, which is
approximately the peak activity seen in late-type stars. Thus we expect
our X-ray selected sample to be biased toward the most
active dMe stars in NGC 2547 and we can see by comparing
Figs.~\ref{vvin2547}a and 1b that there are probably many more members
of NGC 2547 in this colour range with lower X-ray activity.
Furthermore, because the X-ray survey is flux-limited, we also expect
the bulk of any detected field population to have $L_{\rm x}/L_{\rm
bol}\sim10^{-3}$, so in this portion of the CMD, the cluster and field
sources coincide.  We believe that several of the spectroscopic
targets in this colour range could be non-members. Two such objects
(RX\,2 and RX\,98) have been identified from their narrow-band spectral
indices and one (RX\,60) has a discrepant RV.

In summary, our spectroscopic sample, refined to include those objects
with photometry, RV and narrow-band indices consistent with cluster
membership, should be unbiased with respect to X-ray activity and
contamination-free for $1.0<V-I_{\rm c}<1.5$; moderately biased toward
the most X-ray luminous stars but contamination free for $1.5<V-I_{\rm
c}<2.5$; and heavily biased towards the most X-ray active cluster
members and possibly still containing one or two contaminating active
dMe field stars for $V-I_{\rm c}>2.5$.

\section{Lithium in NGC 2547}

\subsection{Lithium abundances: comparison with theoretical models}

\begin{figure}
\vspace*{7.5cm}
\includegraphics{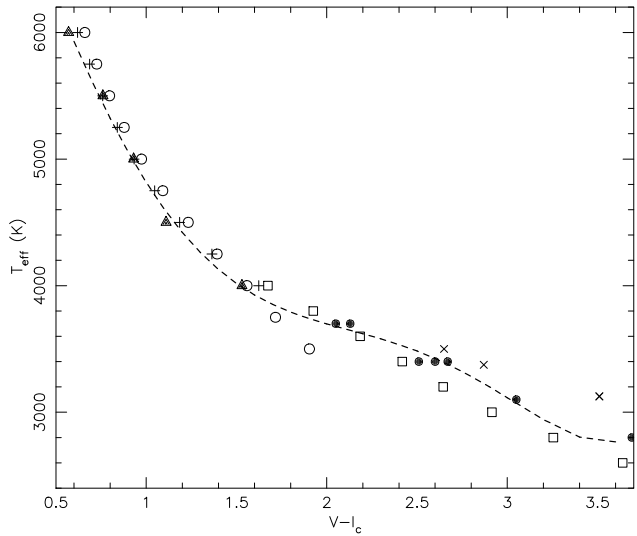}
\caption{The relationship between $T_{\rm eff}$ and intrinsic
$V-I_{\rm c}$ used in this paper is shown as a dashed line. The open
circles and squares are based upon synthetic photometry from the {\sc
atlas9} and {\sc nmarcs} atmospheric models respectively (Bessell et
al. 1998); the crosses are synthetic photometry from the {\sc
marcs-ssg} models (Houdashelt et al. 2000) 
the solid triangles, spots and diagonal crosses are empirically estimated temperatures
from Bessell (1979), Leggett (1996) and Kirkpatrick et al. (1993)
respectively.}
\label{teffviplot}
\end{figure}

\begin{figure}
\vspace*{8.3cm}
\includegraphics{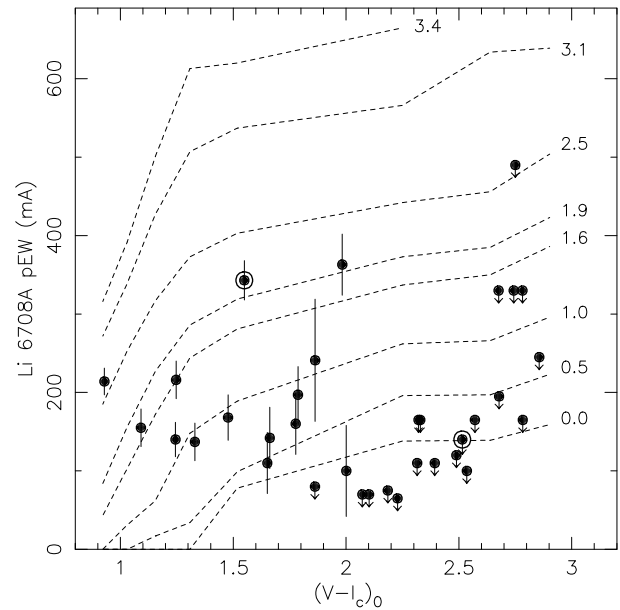}
\caption{Li pEW vs intrinsic $V-I_{\rm c}$ for members of NGC 2547. 
The two circled points are either spectroscopic binaries or cluster
non-members (see Sect.~\ref{member}). The
dashed lines are loci of constant Li abundance (labelled with their
$A$(Li) values) transformed into the
observational plane using our spectral syntheses for $T_{\rm
eff}>4000$\,K, the curves of growth of Zapatero Osorio et al. (2002)
for lower $T_{\rm eff}$ and using the $T_{\rm eff}$-$V-I_{\rm
c}$ relationship plotted in Fig.~\ref{teffviplot}.
}
\label{liewviplot}
\end{figure}

\begin{figure}
\vspace*{7.5cm}
\includegraphics{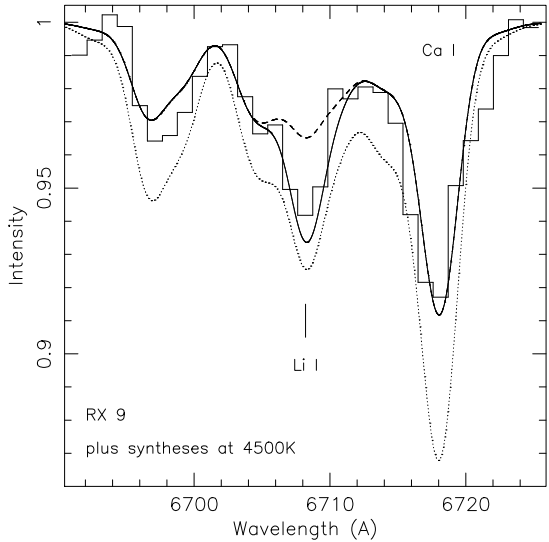}
\caption{The spectrum of RX\,9 around the \lii\ feature. Three
syntheses for $T_{\rm eff}=4500$\,K are shown. Solid line -
[M/H]$=-0.25$ and $A$(Li)$=1.6$, dashed line - [M/H]$=-0.25$ and
$A$(Li)$=1.0$, dotted line - [M/H]$=0.0$ and $A$(Li)$=1.6$. We estimate
RX\,9 has $T_{\rm eff}=4347$\,K so the implication is that the star
has a sub-solar metallicity.}
\label{lisynth}
\end{figure}

Models of PMS Li depletion predict Li abundances as a
function of $T_{\rm eff}$ (or equivalently, mass) and age. To compare
Li in NGC 2547 with models we calculated effective temperatures
and estimated Li abundances with atmospheric syntheses. 

For the conversion between $V-I_{\rm c}$ and $T_{\rm eff}$ we chose an
empirical relationship matching data in Bessell (1979) for $T_{\rm
eff}>4000$\,K and young disk stars from Leggett (1996) at cooler
temperatures (see Fig.~\ref{teffviplot}).  The correct form of these
relationships at low temperatures is still controversial, although the
agreement between empirical measurements and synthetic photometry from
models for $T_{\rm eff}>4000$\,K is encouraging (e.g. Bessell, Castelli
\& Plez 1998; Houdashelt, Bell \&
Sweigart 2000). Synthetic photometry based on model atmospheres would predict hotter
temperatures (by $30-50$\,K) at $V-I_{\rm c}\simeq1.2$, but cooler
temperatures (by $\sim100$\,K at $V-I_{\rm c}\simeq2.6$ (Bessell et
al. 1998; Houdashelt et al. 2000). The $V-I_{\rm c}$-$T_{\rm eff}$ relationship
empirically defined by Kirkpatrick et al. (1993) would give temperatures about
75\,K higher at $V-I_{\rm c}\simeq 2.8$. Beyond this, $T_{\rm eff}$
uncertainties could exceed 200\,K.

For $T_{\rm eff}>4000$\,K we
synthesized spectra with the {\sc uclsyn} code (Smith 1992; Smalley,
Smith \& Dworetsky 2001), which uses {\sc atlas9} atmospheres (Kurucz
1993) incorporating a mixing length theory of convection with
$\alpha=1.25$, but no overshooting. We ``tuned'' our atomic line list
(no molecules were included) by comparison with a high resolution
solar spectral atlas (Kurucz, Furenlid \& Brault 1984). The solar spectrum was
synthesized at $T_{\rm eff}=5777$\,K, $\log g=4.44$, microturbulence,
$\psi=1.5$\kms\ (see Castelli, Gratton \& Kurucz 1997) and with the chemical
abundances of Anders \& Grevesse (1989). Atomic $gf$ values were
altered until line EWs in the synthesis matched the solar atlas.

We synthesised spectra for the NGC 2547 targets using $\log g=4.5$ and
$\psi=1.5$\kms, blurring the result with a Gaussian that matched the
resolution of our spectra (although not the detailed shape of the instrumental
profile, which varied slightly between spectra).
The model spectra did not match the observations very well
unless we reduced the metallicity (Fe, Ca and Al which all have strong
neutral lines in the vicinity of the \lii\ 6708\AA\ doublet) by about
0.25\,dex (see for example Fig.~\ref{lisynth}).  Some of this could be
due to errors in the $T_{\rm eff}$ scale, but the stars would have to
be 300\,K hotter for a solar metallicity to provide a reasonable match.
Naylor et al. (2002) also noted that there was some evidence for a
sub-solar metallicity from a comparison of fits to the $V,B-V$ and
$V,V-I_{\rm c}$ colour-magnitude diagrams. For now, we claim that
NGC 2547 {\em probably} has a sub-solar metallicity, but this needs
confirmation with better quality spectra of solar-type stars. Fixing
the metallicity at [M/H]$=-0.25$, we generated spectra at a grid of
$A($Li$)$ ($= 12 + \log $N(Li)/N(H)) and $T_{\rm eff}$ points and
determined the pEW of the \lii\ 6708\AA\ feature in the same way as we
did for the target spectra.  We adopted this approach rather than
trying to fit models to every spectrum because: 
(a) at cooler temperatures we are reliant on pEW curves of
growth supplied by other authors (see below); (b) we need to apply a
consistent analysis to stars in other clusters where the original
spectra are not published;  (c) the instrumental profile varied
somewhat from fibre to fibre and was not entirely Gaussian.  Adopting
the sub-solar metallicity resulted in predicted pEWs (at a given Li
abundance) that were about 10 per cent larger than for a solar
metallicity, because of the reduced influence of the wings of the strong \cai\
6718\AA\ line on the continuum flux around \lii\ 6708\AA.

Synthetic photometry from the {\sc atlas9} models (open circles in
Fig.~\ref{teffviplot}), upon which we based our spectral
syntheses, diverge from empirical $T_{\rm eff}$-$V-I_{\rm c}$
measurements below 4000\,K. This is not surprising given that these
atmospheres do not include opacity from molecular species such as TiO
and H$_{2}$O which begin to form at these temperatures. They are at
least partially included in the {\sc nmarcs} models, which seem to
follow the empirical data better (see Bessell et al. 1998).  For
$T_{\rm eff}\leq 4000$\,K we have used curves of growth relating the
pEW of the \lii\ 6708\AA\ line, $T_{\rm eff}$ and Li abundance
presented by Zapatero Osorio et al. (2002 -- their Table~5),
supplemented by data for pEWs at lower Li abundances (down to
$A$(Li)$=0.0$) supplied by
Pavlenko \& Zapatero Osorio (private communication).  These were
calculated using synthetic spectra generated from the atmospheric
models of Hauschildt, Allard \& Baron (1999), incorporating atomic and
molecular opacities. Furthermore, the curves of growth were calculated
for pEWs measured at a spectral resolution of 1.7\AA\ -- not too
dissimilar to the data considered here. Unfortunately, apart from the
highest Li abundances, the calculations are performed for $\log g=4.0$,
which is a little too low for our targets. We have experimented using
{\em our} {\sc atlas9} synthetic spectra at 4000\,K and find that
increasing $\log g$ to 4.5 at a given Li abundance results in only a 10
per cent {\em lower} pEW for $A$(Li)$<2.5$ and roughly 10 per cent {\em
higher} at $A$(Li)$>3.0$. This latter result agrees quantitatively with what
Zapatero Osorio et al. found but the predicted pEWs from our models
are larger: -- at 4000\,K, $\log g=4.5$ and $A$(Li)$=3.1$, the {\sc
atlas9} synthesis predicts a pEW of 668\,m\AA\, whereas Zapatero Osorio
et al. give 537\,m\AA. Only a small part of this discrepancy is due to
our differing resolutions and pEW definitions, the rest is due to the
lowering of the continuum by TiO absorption at cooler temperatures (see
Fig.~14 in Zapatero Osorio et al.). We choose simply to apply the
results of the Zapatero Osorio et al.  analysis to our pEWs,
recognising that there may be small systematic errors of order 10 per cent in
the curves of growth.

The curves of growth and data for NGC 2547 are illustrated in
Fig.~\ref{liewviplot}, where $T_{\rm eff}$ values have been transformed to
$V-I_{\rm c}$ using the relationship defined in Fig.~\ref{teffviplot}.
Li abundances for our stars are obtained using bi-cubic spline
interpolation.  Uncertainties in these
abundances have internal and external components.  Internal
uncertainties are a consequence of uncertain $T_{\rm eff}$ and pEW
values. Changes of microturbulence (by 0.5\kms) make a comparatively negligible
difference to derived Li abundances. A change of pEW by 40m\AA\
changes $A$(Li) by 0.1-0.3\,dex depending on $T_{\rm eff}$ and
pEW. The photometry for NGC 2547
has an internal precision of better than 0.04 magnitudes (Naylor et
al. 2002), which translates to a $T_{\rm eff}$ uncertainty of
$\leq50$\,K for the stars in our sample. Photometric variability due to
starspots may mean this is optimistic, although the narrowness of
spread around an isochrone for the single stars (see
Fig.~\ref{vvin2547}b) probably means that uncertainties in $V-I_{\rm
c}$ must be much less than 0.1 magnitude. We conservatively adopt a
relative $T_{\rm eff}$ error of $\pm100$\,K for all our stars.  This
leads to relative Li abundance errors of about 0.2\,dex in the hotter
stars, but much less than this for $T_{\rm eff}<4000$\,K. The uncertainties
in the Li abundances listed in Table~1 result from an individual
calculation for each star based on these factors. 

External uncertainties arise as a
consequence of our assumed $T_{\rm eff}$ scale and from any
deficiencies in the adopted atmospheric models and curves of growth. 
Some idea of the uncertainties in the $T_{\rm eff}$ scale can be gleaned from
Fig.~\ref{teffviplot}. It seems that uncertainties remain at about the
$\pm100$\,K level for $T_{\rm eff}>3800$\,K, perhaps increasing to
$\pm150$\,K at lower temperatures. This only leads to systematic
abundance uncertainties of $\pm0.2$\,dex at most. Related to this is any
systematic calibration problem with the $V-I_{\rm c}$ colours. This
will be small for the hotter stars, but could be as large as 0.1 mag
for $V-I_{\rm c}>2.0$ ($T_{\rm eff}<3700$\,K). However, this leads to
systematic $T_{\rm eff}$ uncertainties smaller than those discussed
above. There is then the remaining systematic uncertainty (for the
cooler stars) due to the adoption of the Zapatero Osorio et al. curves
of growth and the difficulties of the differing spectral resolutions
and adopted $\log g$. We estimate this could result in pEW
predictions for a given abundance being systematically uncertain by around 10 per
cent, which leads to a further systematic abundance uncertainty of about 0.2
dex, roughly independent of temperature between 3000\,K and
4000\,K.

Finally, our curves of growth are for LTE formation of the \lii\
6708\AA\ resonance doublet. Calculations presented by Carlsson et
al. (1994), Pavlenko et al. (1995) and Pavlenko \& Magazzu (1996)
concur that the correction is about 0.2\,dex (in the sense that the
NLTE abundance is smaller) for $T_{\rm eff}\simeq5000$\,K and
$A$(Li)$_{\rm LTE}=3.5$, becomes negligible for $2<A$(Li)$_{\rm LTE}<3$
and then decreases to a minimum of $-0.3$\,dex for $T_{\rm eff}=4500$\,K
and $A$(Li)$_{\rm LTE}\simeq1.0$.  At $T_{\rm eff}\leq 4000$\,K, both
Pavlenko et al. (1995) and Pavlenko \& Magazzu (1996) indicate that the
NLTE corrections become much smaller.  We make no correction to the LTE
abundances listed in Table~1.

\label{liabun}

\subsection{Lithium abundances: comparison with other clusters}

\label{licomp}

Comparing our data to those in other clusters at similar, younger and
older ages provides further empirical constraints on the amount of Li
depletion in the evolutionary models.
\begin{enumerate}
\item Young PMS stars in the Sigma Orionis cluster (Zapatero Osorio et
al. 2002) and the Taurus-Auriga star forming region (Mart\'{\i}n et
al. 1994): We choose only those stars classed as ``weak-lined'' T-Tauri
stars, to avoid any possible complications with a veiling continuum
from an active accretion disc.  The ages of stars in this sample lie in
the range 1-8\,Myr.  The spectral resolution of these
observations were about 1.7\AA\ and 0.2-0.7\AA\ respectively.
\item IC 2391 and IC\,2602: These clusters are approximately co-eval
and have ages (from low-mass isochrone fits) that are slightly older
than NGC 2547 by 5-10\,Myr (Jeffries \& Tolley 1998). A more recent
determination of the age via the ``Lithium depletion boundary method''
(see Barrado y Navascu\'{e}s, Stauffer \& Patten 1999 and Sect.~\ref{discuss})
yields an age of $53\pm5$\,Myr for IC\,2391. Li EWs are given by
Randich et al. (1997, 2002) for an X-ray selected sample of low-mass
stars in IC\,2391 and IC\,2602. Stauffer et al. (1989) present Li EWs
for a small sample of photometrically selected IC\,2391 members. All
these data were obtained at higher spectral resolutions (between 0.15
and 0.5\AA) than our own. $V-I_{\rm c}$ photometry comes
from Patten \& Simon (1996), Prosser, Randich \& Stauffer (1996) or
Stauffer et al. (1989). The latter quote Kron $V-I$ colours
which are converted into $V-I_{\rm c}$ using the formula of Bessell \& Weis (1987).
\item The Pleiades: This cluster has an isochronal age of about
100\,Myr and an age determined from the lithium depletion boundary of
$125\pm8$\,Myr (Stauffer, Schultz \& Kirkpatrick 1998).  Li EWs for
cool Pleiades stars were found in Soderblom et al. (1993);
Garc\'{\i}a-L\'opez et al. (1994) and Jones et al. (1996). These data were
taken at spectral resolutions of between 0.15\AA\ and 0.7\AA. $V-I$
photometry for these stars can be found in the Open Cluster
Database\footnote{The Open~Cluster~Database, provided by C.F. Prosser and
J.R. Stauffer. Accessed at\\ {\tt
ftp://cfa-ftp.harvard.edu/pub/stauffer/clusters/}}
and again, are transformed to $V-I_{\rm c}$. For stars common to these
literature sources, we followed the same precedence order described in
detail by Randich et al. (2001).
\end{enumerate}

For each dataset we determined LTE Li abundances from the published
6708\AA\ pEWs using methods similar to those described in
Section~\ref{liabun}. $T_{\rm eff}$ for the Pleiades and IC\,2391/2602
stars was estimated from the $V-I_{\rm c}$ photometry, using the same
colour-$T_{\rm eff}$ relation as for NGC 2547. We used $E(V-I_{\rm
c})=0.05$ for the Pleiades and 0.01 and 0.04 for IC\,2391 and IC\,2602
respectively (Patten \& Simon 1996). The same relationship was {\it
not} used for the younger PMS stars, where $V-I_{\rm c}$ was not
available. Instead, we used $T_{\rm eff}$ values quoted in Mart\'{\i}n
et al. (1994) and Zapatero Osorio et al. (2002), which come from the
spectral types and relationships proposed by de Jager \& Nieuwenhuijzen
(1987) and Bessell (1991) respectively. Whilst there are likely zero
point errors of order 100-150\,K associated with these choices (see
Mart\'{\i}n et al. 1994 for discussion), such uncertainties lead only to small
Li abundance errors in cooler stars (see Sect.~\ref{liabun}), which are
inconsequential for this paper.

To account for the differing spectral resolutions of the various
datasets (and also differing amounts of rotational broadening from
star to star), we generated ten (solar metallicity) curves of growth for the \lii\
6708\AA\ feature blurred with Gaussians to simulate FWHMs of 0.15\AA\ to
2.7\AA\ and with $5250\leq T_{\rm eff} \leq 4000$\,K and $\log g$ of 4.0
and 4.5. Bicubic spline interpolation was used to find the LTE Li
abundance for each star, using the curve of growth most appropriate for
the instrumental resolution and rotational broadening ($v \sin i$
values are available from the same literature sources as the \lii\
pEWs). We used the $\log g=4.0$ curves of growth for the very young PMS
stars and the $\log g=4.5$ curves of growth for IC\,2391/2602 and the
Pleiades.

For $T_{\rm eff}<4000$\,K we are forced to use the curves of growth
supplied by Zapatero et al. (2002) despite these being strictly
appropriate only for the data taken at lower resolutions (or fast
rotating stars with $v \sin i\ga 50$\kms). However, the difference
between the predicted pEWs at different resolutions is only significant
($>10$ per cent) when the pEWs become smaller than about 0.2\AA\
(corresponding to $A$(Li)$<0.5$ at these temperatures). This only
affects a very few cool stars in the Pleiades and IC\,2391/2602 and for
these stars we will have {\em overestimated} the Li abundances, because the
effect of poorer resolution is to reduce the observed pEW of weaker
\lii\ lines.

\section{Discussion}

\label{discuss}

\subsection{Comparison with theoretical models}
\label{theory}

\begin{figure*}
\vspace*{20cm}
\includegraphics{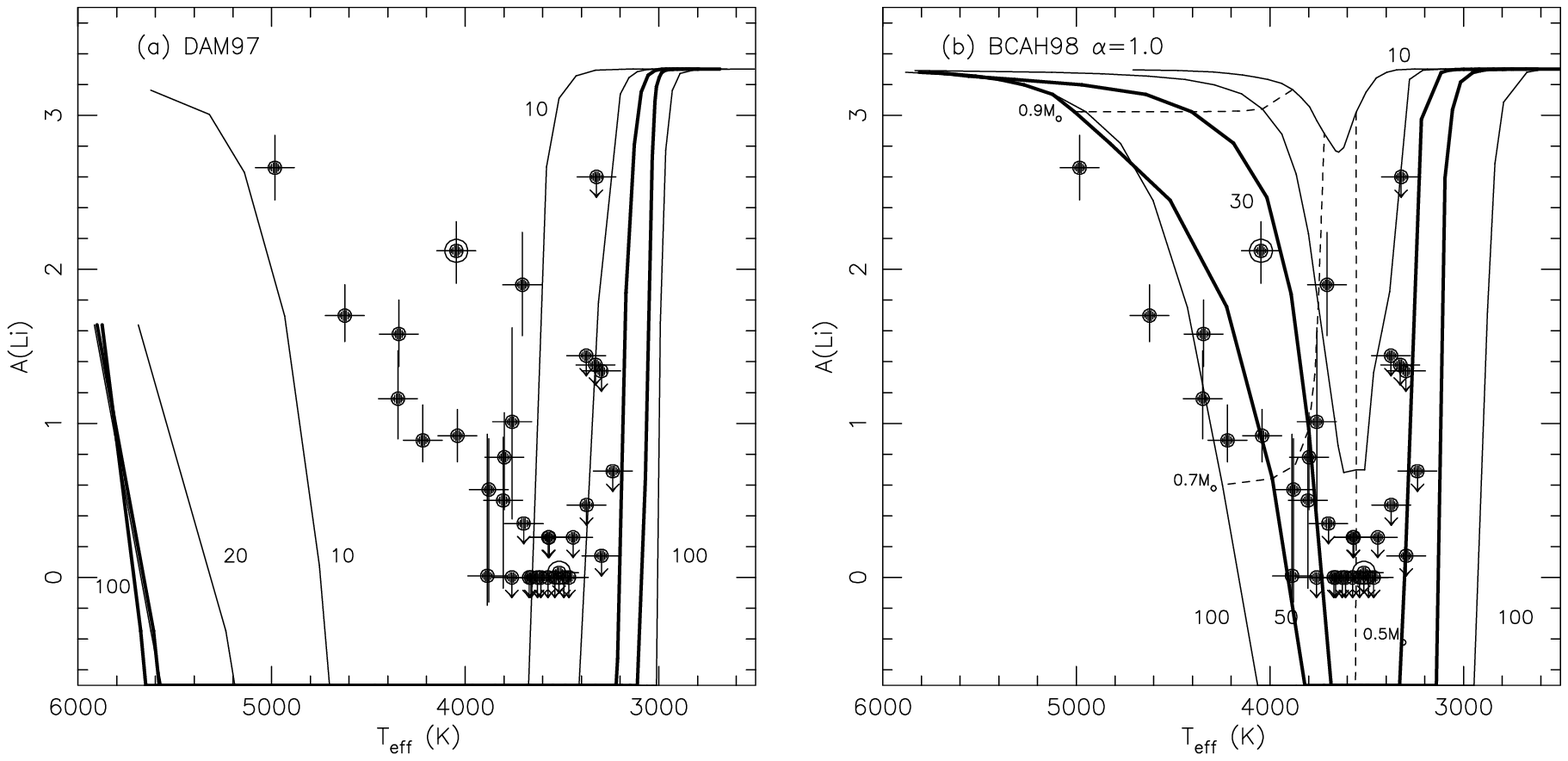}
\includegraphics{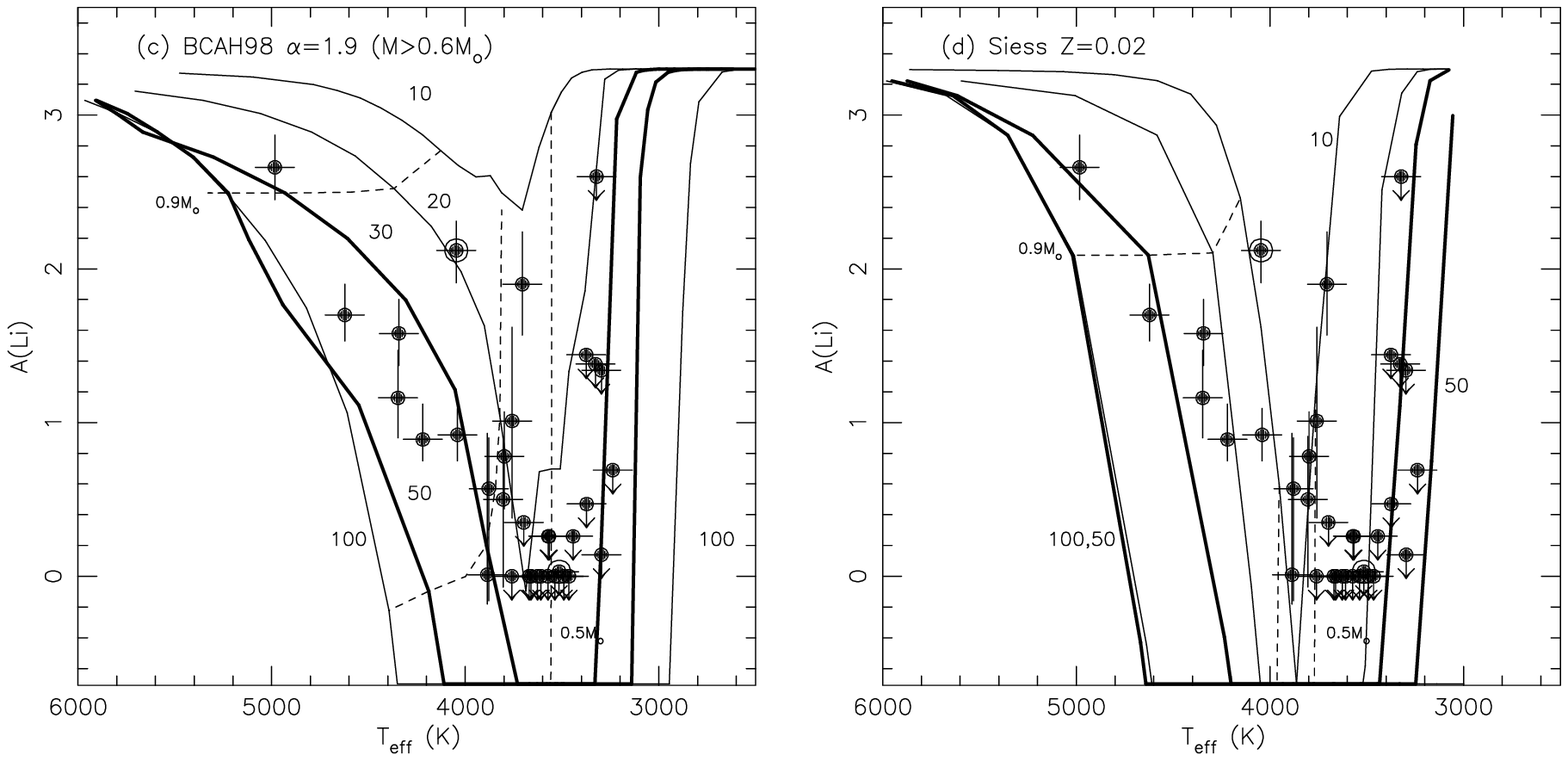}
\includegraphics{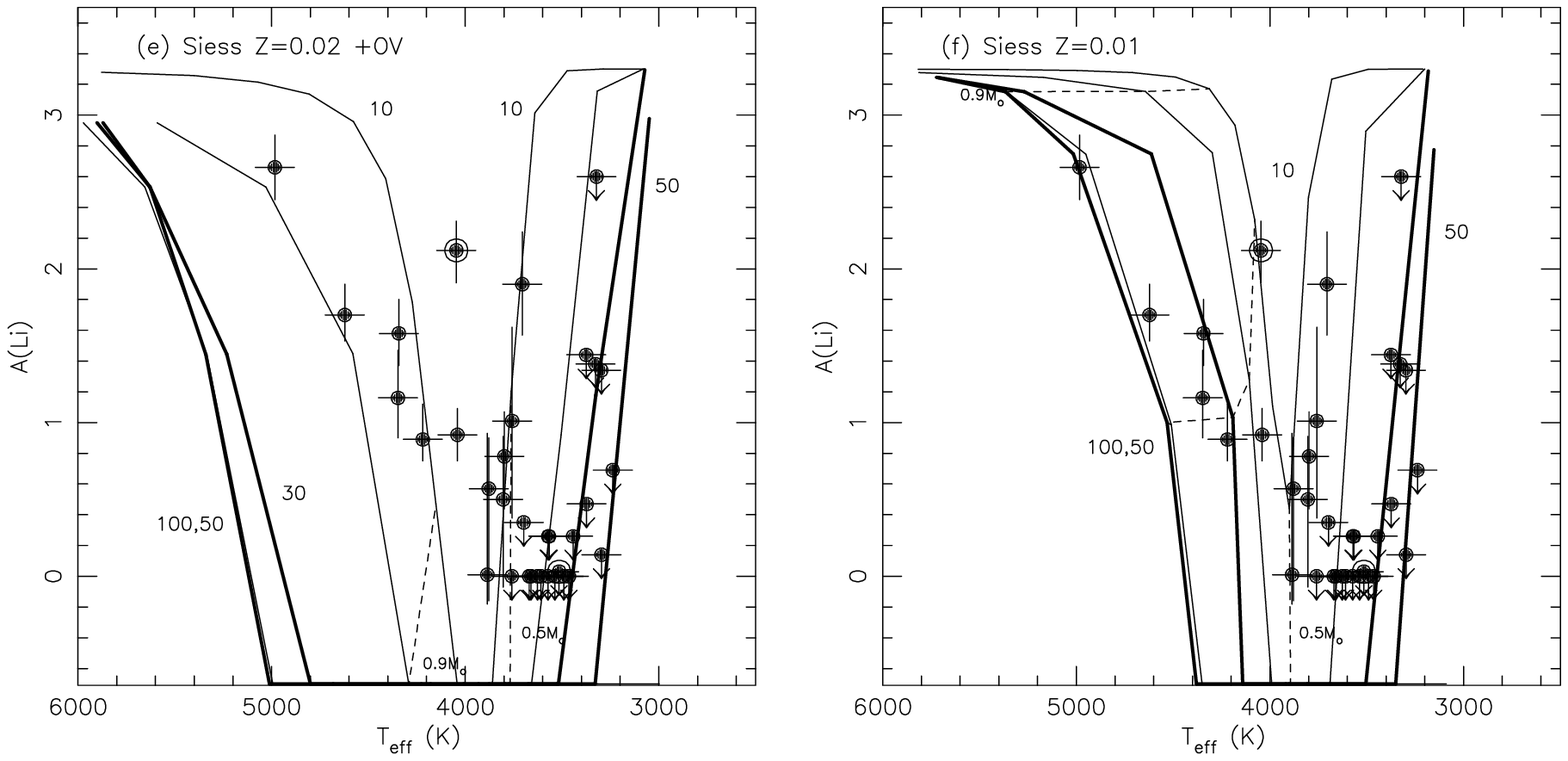}
\caption{Comparison of the NGC 2547 LTE Li abundances with theoretical
models. The two circled points are either spectroscopic binaries or
cluster non-members (see Sect.~\ref{member}). 
(a) D'Antona \& Mazzitelli (1997), (b) Baraffe et al. (1998)
with a mixing length parameter of 1.0 pressure scale-heights, (c)
Baraffe et al. (1998, 2002) with the mixing length parameter set to 1.9
pressure scale-heights for $M>0.6M_{\odot}$, (d) Siess et al. (2000)
with $Z=0.02$ and no overshoot, (e) Siess et al. (2000) with $Z=0.02$ and 0.2 pressure
scale-heights of overshoot, (f) Siess et al. (2000) with $Z=0.01$ and no
overshoot. For each set of models we show 10, 20, 30 (thicker line), 50
(thicker line) and 100\,Myr
isochrones of Li depletion (assuming an initial $A$(Li)$=3.3$). Where
possible we show (as dashed lines) evolutionary tracks for stars
of mass 0.5, 0.7 and 0.9$M_{\odot}$. The error bars on the NGC
2547 points arise from random $T_{\rm eff}$ and pEW
errors. Additional systematic errors are discussed in Sect.~\ref{liabun}.
}
\label{modelcomp}
\end{figure*}

Armed with Li abundances we can compare them with the predictions of
theory. In this paper we consider three groups of models with readily
available Li depletion factors as a function of $T_{\rm eff}$ and age
(D'Antona \& Mazzitelli 1997; Siess, Forestini \& Dufour 2000; Baraffe
et al. 1998, 2002).  We assume that the {\em initial} Li abundance is
similar to that in meteorites and very young T-Tauri stars --
$A$(Li)\,$=3.3$. There is some latitude for uncertainty in this, although
the uncertainty is probably less than the systematic abundance errors
discussed in Sect.~\ref{liabun}.

We show the comparisons in the six panels of Fig.~\ref{modelcomp}, where
Li abundances are plotted on top of isochrones in the $A$(Li)-$T_{\rm
eff}$ plane. Of the systematic abundance errors
mentioned in Sect.~\ref{liabun}, it is likely that only the upward NLTE
corrections of $\leq 0.3$\,dex at around 4500\,K and the 0.2\,dex curve
of growth uncertainties for $T_{\rm eff}<4000$\,K are
important. A systematic $T_{\rm eff}$ scale uncertainty is less
important -- $T_{\rm eff}$ and $A$(Li) will change in such a way that
points tend to move {\em along} the trends defined by the models and the data
for $T_{\rm eff}>3800$\,K.

The models highlight the sensitivity of predicted
Li abundances to various physical parameters and approximations.
Figure~\ref{modelcomp}a shows isochrones from the $Z=0.018$ D'Antona \&
Mazzitelli (1997) models, featuring a grey atmosphere approximation and
full spectrum turbulence (FST) model for convection.  Figures~\ref{modelcomp}b
and c show $Z=0.018$ isochrones from the models described by Baraffe
et al. (1998, 2002).  These have a detailed atmosphere (including
contributions from molecular opacity -- Hauschildt et al. 1999) 
and convective energy transport
using mixing length theory (MLT).  Figure~\ref{modelcomp}b sets the mixing
length parameter to 1.0 pressure scale height, whereas in
\ref{modelcomp}c 
it is set
to 1.9 pressure scale heights for masses above 0.6$M_{\odot}$. The
isochrones in Figs.~\ref{modelcomp}d, e and f are from Siess et
al. (2000). These use MLT with the mixing length
parameter set to 1.6 pressure scale heights. The atmosphere is
constructed from ``analytic fits'' to the predictions of detailed
atmosphere models, as a function of $T_{\rm eff}$, gravity and
metallicity. Figure~\ref{modelcomp}d shows the basic models with
$Z=0.02$, Fig.~\ref{modelcomp}e the effect of adding 0.2 pressure scale
heights of overshoot at the convective boundaries and
Fig.~\ref{modelcomp}f shows the effect of decreasing the metallicity to
$Z=0.01$.

A detailed discussion of the differences between the models is given
by Siess et al. (2000) and Baraffe et al. (2002).  The most
important thing to say about Fig.~\ref{modelcomp}, is that it
illustrates that Li abundances {\em should} be used as critical tests
of these models. The variation in predicted Li abundance is enormous --
at 30\,Myr and 4500\,K, there are several orders of magnitude of
difference in the Li-depletion predictions.  These differences are much
larger than the systematic errors in measuring $A$(Li) and $T_{\rm
eff}$ and much more dramatic than predicted differences between
isochrones in the Hertzsprung-Russell diagram. The comparison also does
not require a precise distance to the cluster.  

Following arguments presented by Siess et al. (2000) and Baraffe et
al. (2002), we surmise that there are two main effects that drive
discrepancies in the predictions of Li depletion by models. Each is
important on one side of the ``Li-chasm'' centered at $\simeq 3800$\,K: (i) the
treatment of convection in stars with $M>0.6M_{\odot}$ where Li
depletion can be halted as a result of an expanding radiative core;
(ii) the treatment of stellar atmospheres for $T_{\rm eff}<4000$\,K, as
molecules become prominent. In other words, the detailed treatment of
convection is not very important in the lowest mass, fully convective
stars and a grey body approximation works reasonably well at higher
temperatures.

We see from Fig.~\ref{modelcomp}a that the D'Antona \& Mazzitelli
(1997) models entirely fail to match observed Li abundances in NGC
2547. The FST convection treatment produces far
too much Li depletion for $T_{\rm eff}>3700$\,K. The {\em same} models
give an isochronal age of about $25\pm5$\,Myr (the uncertainty arises mainly
from the distance) for NGC 2547 in the $V$
vs $V-I_{\rm c}$ CMD using the same stars (Naylor et al. 2002), so
there is a clear inconsistency.  Ventura et al. (1998) and D'Antona,
Ventura \& Mazzitelli (2000) have suggested that dynamo generated
magnetic fields might provide an increase in the temperature gradient
required for convective instability, decreasing temperatures at the
convection zone base and drastically reducing the predicted Li
depletion in solar-type stars (discussed  further in Sect.~6.3). 

The Baraffe et al. (1998, 2002) models fare better.
These models also give isochronal ages in the
$V,V-I_{\rm c}$ CMD of 25 to 35\,Myr (with the older age favoured by
the smaller mixing length model).
The models in Fig.~\ref{modelcomp}b do not deplete enough Li at
$4000<T_{\rm eff}<5000$\,K (even taking into account NLTE corrections), 
although at slightly cooler temperatures
there is a reasonable match to the 30\,Myr isochrone. (RX\,22A  and the
possible cluster spectroscopic binary RX\,76A
have exceptionally high Li abundances and are discussed in Sect.~6.4)
Additional Li depletion might be possible if
there is extra mixing in the radiative layers beneath the
convection zone base. Overshooting is an example of this and
comparison between Figs.~\ref{modelcomp}d (without overshooting) and
\ref{modelcomp}e (with moderate overshooting) shows that this could be
very important in the hotter stars. Other possibilities, reviewed by
Pinsonneault (1997), include slow mixing due to turbulence induced
by differential rotation or gravity waves. However, these are likely to
act on {\em much} longer timescales than the age of NGC 2547
(Chaboyer et al. 1995).  Figure~\ref{modelcomp}c shows
that additional mixing may not be needed. These models, featuring a
larger mixing length parameter for $M>0.6M_{\odot}$ (and which are
found to give a much better match to the solar structure), 
fit the NGC 2547 data reasonably well over the whole $T_{\rm
eff}$ range at an age of $20-30$\,Myr
(given the systematic uncertainties in $T_{\rm eff}$ and
$A$(Li), especially the likelihood that the LTE abundances should be
increased by 0.2-0.3\,dex at $\sim$4500\,K), with the exception of the
high point RX\,22A.

The Siess et al. (2000) $A$(Li)-$T_{\rm eff}$ isochrones in
Figs.~\ref{modelcomp}d,e,f demonstrate that Li depletion is sensitive
to overshoot (see above) and metallicity. Reducing the metallicity
should result in less Li depletion at a given $T_{\rm eff}$ because the
opacities are lower and convection zone bases cooler in stars with
central radiative zones. In addition the metal-poor stars are hotter at a
given mass by 100-200\,K, which accounts for the majority of the
difference between the Li depletion isochrones on the cool side of the
``Li-chasm''. The models with $Z=0.02$ and no overshoot have problems at
around 3700-4000\,K, where they predict that Li should be undetectable
in NGC 2547. The models with overshoot make this problem worse. The
$Z=0.01$ models are a better match to the data at 4000\,K, although
RX\,22A and RX\,76A remain unexplained. Note that all the Siess et
al. models have very little evolution of the Li depletion pattern on
the hot side of the Li-chasm between 50 and 100\,Myr -- presumably
because the convection zone base has become too cool to burn Li and
the star has either reached or is dropping vertically onto the ZAMS in
the Hertzsprung-Russell diagram.

In principle the region on the cool side of the Li chasm can also be
used to test evolutionary models.  Stars show an abrupt
depletion of their surface Li when the core becomes hot enough to
initiate Li burning -- the lithium depletion boundary (LDB).  The
$T_{\rm eff}$ at which this occurs is a sensitive function of age and
an age derived in this way should agree with that from isochronal fits
to the low-mass stars in the Hertzsprung-Russell diagram {\em and} the
amount of Li depletion seen on the hot side of the Li chasm.  As we
have no Li detections on the cool side of the Li chasm, we can only put
lower limits on the LDB age.  Figure~\ref{modelcomp} suggest lower
limits to the age of 20$-50$\,Myr. In fact, because the drop in Li
abundance on the cool side of the ``Li-chasm'' is so sharp and the
$T_{\rm eff}$ scale so uncertain, the comparison is best done against
bolometric luminosity.  This question is examined in much greater
detail by Oliveira et al. (2003) using a later 2dF dataset of an even
fainter sample of NGC 2547 candidates. They find, using the same set of
evolutionary models, that the LDB age is
definitely older than 35\,Myr, but probably younger than 54\,Myr.

\subsection{Comparison with other clusters}

\begin{figure*}
\vspace*{8cm}
\includegraphics{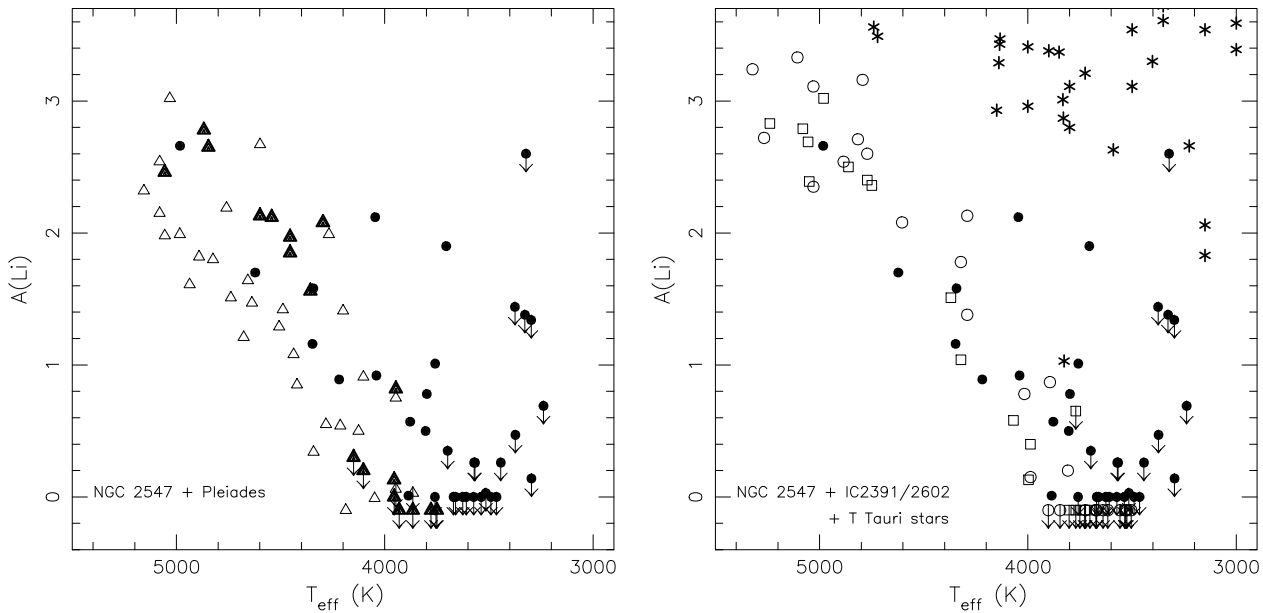}
\caption{A comparison of LTE Li abundances among the cool stars of: (a) NGC
2547 (spots) and the Pleiades (triangles). Rapid rotators in the
Pleiades ($v \sin i > 20$\kms) are represented by large, filled
triangles; (b) NGC 2547 (spots), IC\,2391 (squares), IC\,2602 (open
circles) and 1-8\,Myr weak-lined T-Tauri stars (stars).}
\label{clustercomp}
\end{figure*}

Figure~\ref{clustercomp} shows the Li abundances in NGC 2547 compared
with those in the Pleiades, IC\,2391/2602 and a sample of (1$-$8)\,Myr
weak-lined T-Tauri stars (see Sect.~\ref{licomp}). The bulk of the
Pleiades stars lie below the NGC 2547 sample by $\sim 0.3$\,dex and
there are detections of Li in NGC 2547 at $T_{\rm eff}$ values slightly
cooler than the point at which Li becomes undetectable in the Pleiades.
However, it is well known that
a relationship exists between measured Li EW (and hence deduced
Li abundance) and rotation in the Pleiades and
other young clusters (e.g. Soderblom et al. 1993).  
This is not a one-to-one mapping, but for stars
with $5500<T_{\rm eff}<4500$\,K it seems that fast rotating stars
always have relatively high Li abundances (compared to their siblings
at similar $T_{\rm eff}$), while slower rotators can have either high
or low Li abundances (Randich et al. 1998; Jeffries 2000). It is claimed that this
relationship is reduced or disappears for $T_{\rm eff}<4500$\,K
(Garc\'{\i}a-L\'opez et al. 1994; Jones et al. 1996).

Whether this trend truly reflects differences in the Li abundances or
rather deficiencies in our understanding of the atmospheres of rapidly
rotating and hence magnetically active stars, is still the subject of
debate (see Barrado y Navascu\'{e}s et al. 2001; Ford,
Jeffries \& Smalley 2002). Nevertheless, we see the same phenomenology in our derived
Pleiades Li abundances (Fig.~\ref{clustercomp}a) and looking at the
rapid rotators only ($v \sin i >20$\kms\ -- though note that some fast
rotators could still be present in the rest of the sample if a small
$\sin i$ allows them to masquerade as slow rotators), it seems
they have a nearly indistinguishable Li depletion pattern
from NGC 2547 and there may even be rapid rotators with {\em less} Li
depletion, though the sample size in NGC 2547 is small.  Now whilst it
is possible that the NGC 2547 sample consists only of similar rapid
rotators the weight of evidence is that this is not the case. In
Sect.~\ref{contaminate} we showed that the NGC 2547 sample is unlikely
to be biased with respect to activity and rotation for $V-I_{\rm
c}<1.5$ ($T_{\rm eff}\ga 4100$\,K) and not heavily biased until much
lower temperatures. Additionally, Randich et al. (1997, 2001) have
shown that cool K and M stars in the similarly aged IC\,2391 and
IC\,2602 clusters have a wide spread in rotation rates, although higher
on average than the Pleiades, and Jeffries et al. (2000) have shown
that a similar pattern exists in the G and early K stars of NGC
2547. We therefore conclude, that cool, slowly rotating Pleiads have
depleted a little more Li than their NGC 2547 counterparts, but fast
rotating Pleiads {\it appear} to have been depleted by a similar amount.

It is also worth noting that the Pleiades is not unusual in any
respect. Li abundances have also been determined in NGC 2516 (Jeffries,
James \& Thurston 1998), Blanco 1 (Jeffries \& James 1999) and M35
(Barrado y Navascu\'{e}s et al. 2001), which have similar ages or are a
little older than the Pleiades. Whilst only G and K stars (with $T_{\rm
eff}\ga 4300$\,K) are studied in these papers, they clearly show (see
the comparisons in each of the cited papers) Li abundance patterns very
close to that of the Pleiades, with the lower envelope defined by
predominantly slowly rotating stars and fast rotating stars having
generally higher Li abundances.

Figure~\ref{clustercomp}b empirically demonstrates the effectiveness of
PMS Li-depletion during the first $\sim30$\,Myr of evolution. More than
two orders of magnitude of Li-depletion has occurred between the young
T-Tauri stars and all three $\sim30-50$\,Myr open clusters. The
Li-depletion pattern in NGC 2547 is very similar to that in
IC\,2391/2602. There are a few Li detections in NGC 2547 at $T_{\rm
eff}\simeq3800$\,K with larger abundances than seen in either of the
other clusters, but the numbers of stars involved are small. The IC\,2391/2602
samples are also X-ray selected (see Randich et al. 2001) and
subject to a very similar level of selection bias as a function of
colour. So, the conclusion that the depletion patterns in these three
clusters are very similar seems secure. Randich et al. (2001) report
that for IC\,2602, where they have enough stars in their sample to do
the test, that both the scatter and dependence of Li abundance on
rotation rate is much reduced compared with the Pleiades.
The overall picture we have is that on average, slow rotators in the younger
clusters appear to deplete some Li between their ages and that of the
Pleiades, whereas the Li abundances of the more rapid rotators remain
roughly constant.

Looking at the models in Fig.~\ref{modelcomp} then such behaviour is
difficult to explain if the ages of NGC~2547, IC\,2391 and IC\,2602
are about 30\,Myr. Only the D'Antona \& Mazzitelli (1997) model and Siess et
al. (2000) model with overshoot predict little depletion between 30 and
100\,Myr, but both of these predict far too much Li depletion at
30\,Myr to match the observational data (see Sect.~\ref{theory}).
A better match to the {\it progression} of Li depletion would be
obtained if these clusters were $\sim 50$\,Myr old. At this age the
convection zone base has pushed out to temperatures too low to burn Li
and the stars are at or nearly at the ZAMS, so there is no great change
in either the Li abundance or the $T_{\rm eff}$ between 50 and
100\,Myr. Unfortunately, all the models predict too much Li depletion
at an age of 50\,Myr apart from the Baraffe et al. (1998) models with
a mixing length of 1.0 pressure scale height (Fig.~\ref{modelcomp}b).

\subsection{The age scale of young clusters}

There are now several techniques for estimating the age of an open
cluster. Traditionally, one uses the nuclear turn-off age, which has
been estimated as $55\pm25$\,Myr (Jeffries \& Tolley 1998) for NGC
2547, 35\,Myr for IC\,2391/2602 and 70-100\,Myr for the Pleiades using
models with a small amount of convective core overshoot
(e.g. Mermilliod 1981).  Isochrone fits to low-mass PMS stars can also
give an age estimate. This technique is
currently not especially useful in the Pleiades, where only very
low-mass stars with quite uncertain colours are still descending
towards the ZAMS, but the positions of $0.3\la M \la 1.2M_{\odot}$
stars in IC\,2391/2602 in optical and near IR CMDs
are consistent with ages of 25$-40$\,Myr (Stauffer et al. 1997; Oliveira et al. 2003;
Barrado y Navascu\'{e}s \& Stauffer 2003), while ages of 20$-$35\,Myr
have been derived for NGC\,2547 using identical techniques (Naylor et
al. 2002; Oliveira et al. 2003).

The low-mass isochronal ages are significantly lower than the LDB ages
derived for the younger clusters. Barrado y Navascu\'{e}s et al. (1999)
find $53\pm5$\, Myr for IC\,2391 and Oliveira et al. (2003) find
35--54\,Myr for NGC 2547. For IC\,2391 and the Pleiades
(with an LDB age of $125\pm8$\,Myr -- Stauffer et al. 1998), the LDB
ages are also older than the ``traditional'' nuclear turn-off ages. The
latter result may imply that a modest amount of convective core
overshoot is required in order to increase the nuclear turn-off
ages. The rationale for this conclusion is that the uncertain physics
in the very low-mass stellar models (atmospheres, convection theory)
has a smaller effect on the determined LDB age than that in the high-mass
turn-off stars (convective overshoot, rotation, semi-convection --
e.g. Meynet \& Maeder 1997). This is not a universal view. Song,
Bessell \& Zuckerman (2002) argue that Li burning needs to be hastened to match the
kinematic and isochronal age in a young, PMS binary system, hence decreasing the LDB
ages.

If we assume that the LDB ages {\it are} accurate then the discrepancy
with the low-mass isochronal ages in NGC\,2547 and IC\,2391 must be
addressed. There are two classes of solution. First, that the
colour-$T_{\rm eff}$ relations used to produce isochrones in terms of
observable quantities are strongly gravity dependent. So even though
the isochrones are calibrated to fit the Pleiades at an age of
100$-$130\,Myr, the calibration changes sufficiently at lower gravities
to underestimate the age of the younger clusters. This seems unlikely
because (i) the gravities do not change very significantly ($\simeq
0.2$\,dex) between these ages (ii) the current model atmospheres do not predict any great
change in the colour-$T_{\rm eff}$ relation over this age and mass
range (iii) ages from various optical and near-IR CMDs agree.
A more interesting possibility is that the evolutionary
models for the $0.3\la M \la 1.2M_{\odot}$ stars used to define the PMS
isochronal ages are not correct.

The Li abundances we have derived in this paper offer support for this
second point of view. Both the Baraffe (with mixing length of 1.9
pressure scale heights) et al. (1998, 2002) and Siess et al. (2000,
without overshoot) models do a reasonable job of fitting the Li data at
or around the PMS isochronal age of $\simeq 30$\,Myr, albeit with some
obvious exceptions (see Sect.~6.4). But neither of these models also
predicts the very small amount of depletion seen between the ages of
NGC\,2547, IC\,2391/2602 and the Pleiades. All of the models we have
considered suggest that strong Li depletion continues until an age of
$\simeq 50$\,Myr, supporting the LDB ages.

Both rotation and internal magnetic fields are capable of altering the
structure and Li depletion in low-mass stars and both are physically
well motivated: it is well known that young PMS stars are magnetically
active, have starspots and can be rapidly rotating. These effects are
explored in some detail by Oliveira et al. (2003). Here we only note
that modelling of these phenomena is at an early stage. Pinsonneault et
al. (1998) observationally demonstrate that rapid rotators lie 0.1 mag
(in $V$) above their slower rotating siblings in a CMD (as a result of
having a lower $T_{\rm eff}$ at the same mass). An isochronal age would
therefore be understimated, but only by about 5\,Myr for stars in
NGC\,2547. Mendes et al. (1999) have shown that such rapid rotation
also has a negligible effect on the $A$(Li)-$T_{\rm eff}$ isochrones at
$\simeq 50$\,Myr. 

Stauffer et al. (2003) point out that the spectral
energy distributions of young low-mass stars can be considerably
altered by starspots (caused by surface magnetic field -- see
below). Potentially, this is a threat to the security of ages
determined from fitting isochrones in CMDs. However, we have calibrated
our colour-$T_{\rm eff}$ relationship using the Pleiades as a template,
which has similar levels of chromospheric and coronal magnetic activity
to NGC 2547 (Jeffries \& Tolley 1998; Jeffries et al. 2000). In any
case, Stauffer et al. also show that $V$ vs $V-I_{\rm c}$ CMDs are
likely to be least affected by this phenomenon and that the Pleiades
adequately defines a ZAMS in this CMD.
 
The structural effects of dynamo-generated B-fields may be more
significant.  Preliminary calculations by D'Antona et al. (2000) have
shown that quite plausible B-fields could decrease the $T_{\rm eff}$ of
a 0.95\,$M_{\odot}$ PMS star by as much as 500\,K. Such an extreme
deviation is enough to place a ZAMS star on a 30\,Myr isochrone -- thus
an older cluster could look much younger when the age is determined
from low-mass isochrones.  As mentioned in Sect.~6.1, the presence of
B-field could also significantly reduce the amount of Li depletion at a
given $T_{\rm eff}$; an older cluster would then also appear much
younger if interpreting the Li depletion of stars on the hot side of
the Li-chasm using evolutionary models with no B-field.  At present
there are no detailed published studies of the effects of rapid
rotation and B-fields on the LDB ages, though they are likely to small
(Burke \& Pinsonneault 2000; F. D'Antona -- private communication).

\subsection{Anomalously lithium rich pre main sequence stars}

In this paper we have identified RX\,22A and also perhaps RX\,76A as
objects which have suffered considerably less Li depletion than their
siblings with similar masses.  RX\,22A is classified as a photometric
binary and has a RV in agreement with the cluster mean. RX\,76A is a
photometric binary with an anomalous RV.  One interpretation of these
facts is that neither is actually a member of NGC 2547.  A
more interesting possibility is that both are binary systems and
that for some reason their binarity has resulted in less
Li depletion. We note that IC\,2391, the Pleiades and even older
clusters such as the Hyades possess similar examples of anomalously Li-rich
cool stars (Thorburn et al. 1993; Barrado y Navascu\'{e}s \& Stauffer 1996,
2003; Oppenheimer et al. 1997).

In older clusters these Li-rich stars have been established as short
period binary systems. It is hypothesised that they suffer
less Li depletion because tidal locking
suppresses additional (non-convective) mixing caused by differential
rotation (e.g. Zahn 1994; Ryan \& Deliyannis 1995). In the younger
clusters this hardly seems a likely explanation. The non-standard
mixing mechanisms are probably not very important during PMS evolution.
{\it If} it can be established that the anomalous stars seen in NGC 2547, IC
2391 and the Pleiades are indeed short period binary members of these
clusters, as opposed to say very young interlopers\footnote{At least in the case
of the two anomalously Li-rich low mass Pleiades candidates, it seems that
they are too bright to be cluster binaries and may in fact be interlopers
from a very young stellar association -- Oppenheimer et al. 1997.},
this would instead strongly
argue for a mechanism which {\em inhibits} PMS Li depletion in close
binary systems and which may {\em also} be responsible for the lack of Li
depletion in the anomalous binary systems of the older clusters.  Such
a mechanism could be linked to either rotation or magnetic fields (as
discussed in Sect.~6.3), because PMS binary systems that are
tidally locked, will tend to rotate faster than their single
counterparts.

\section{Summary}

Starting with X-ray selected candidate low-mass ($0.3\leq M \leq
0.9$\,$M_{\odot}$) members of NGC 2547, we have produced a filtered list
of stars which have photometry, spectral indices and radial velocities
consistent with cluster membership. In Sect.~\ref{contaminate} we
showed that this list is unlikely to be complete or uncontaminated and
provide estimates of the size of these problems.

We have calculated LTE Li abundances for the members of NGC 2547
and made a thorough estimate of the likely random and systematic errors
involved in these estimates. The Li abundances have been compared with
theoretical evolutionary models and with other published observations
of Li abundances in younger and older clusters and associations.
\begin{enumerate}
\item We find that some models (specifically those of Baraffe et al. (2002)
featuring a mixing length of 1.9 pressure scale heights for
$M>0.6M_{\odot}$ and those of Siess et al. (2000) with no convective
overshoot) provide a reasonable match to the Li depletion pattern for
$5000<T_{\rm eff}<3700$\,K at the ages of 20$-$35\,Myr which are indicated
by low-mass isochrone fits in the Hertzsprung-Russell diagram 
using the same models. Other models predict
far more Li depletion than seen.
\item However, the models that give a reasonable description of the
Li depletion pattern in NGC 2547 {\it also} predict significant Li depletion between the
ages of 30 and $\simeq 100$\,Myr, disagreeing with observed
$A$(Li)-$T_{\rm eff}$ patterns in the young NGC
2547 and IC 2391/2602 clusters compared with the older Pleiades cluster.
\item The observed {\it progression} of Li depletion in these clusters
rather suggests that NGC 2547 and IC 2391/2602 have ages $\geq
50$\,Myr, which agrees with estimates based on the location of the
``Lithium Depletion Boundary'' seen in even cooler, lower mass stars in
these clusters (Barrado y Navascu\'{e}s et al. 1999; Barrado y Navascu\'{e}s \&
Stauffer 2003; Oliveira et al. 2003).
\item The ages estimated from these techniques, and also from the
nuclear turn-off in high-mass stars, are affected differently by the
physical ingredients that go into the models. We speculate that the
addition of rotation or magnetic fields, both of which are capable of
altering the internal structure of low-mass PMS stars and which are
physically well motivated, may provide a reconciliation of all the age
estimates. Detailed theoretical models and isochrones that incorporate
these effects are called for.
\item We have identified a couple of anomalously Li-rich, low-mass
stars in NGC 2547 that have similarities to objects found in other
young clusters. They may both be short period binary systems in NGC 2547, although
more observational proof is required. If so, this may point to some mechanism
that inhibits the normal process of PMS Li depletion in tidally locked binary systems.
\end{enumerate}

\section*{Acknowledgements}  

We thank the director and staff of the Anglo-Australian Observatory and
especially Russell Cannon who performed the 2dF observations. Thanks
are also due to Nigel Hambly who provided accurate astrometry for our
targets. Computational work was performed on the Keele node of the
PPARC funded Starlink network.  JMO acknowledges the financial support
of the UK Particle Physics and Astronomy Research Council.

\nocite{houdashelt00}
\nocite{baraffe98}
\nocite{baraffe02}
\nocite{leggett92}
\nocite{hambly01}
\nocite{jeffriesblanco199}
\nocite{soderblom93pleiadesli}
\nocite{thorburn93}    
\nocite{pinsonneault97}
\nocite{ventura98}
\nocite{mendes99}
\nocite{chaboyer95}
\nocite{dantona00}
\nocite{ford01}
\nocite{jeffries2000lirev} 
\nocite{naylor98}
\nocite{naylor02}
\nocite{kurucz84}
\nocite{castelli99}
\nocite{bessell98}
\nocite{kurucz93atlas}
\nocite{siess00}
\nocite{stauffer98}
\nocite{carlsson94}
\nocite{zapatero02}
\nocite{leggett96}
\nocite{bessell87}
\nocite{bessell79}
\nocite{kirkpatrick93}
\nocite{patten96}
\nocite{prosseric260296}
\nocite{randich01}
\nocite{randich97}
\nocite{jones96pleiades}
\nocite{garcia94}
\nocite{anders89}
\nocite{barrado99}
\nocite{jeffries98n2547}
\nocite{jeffries00}
\nocite{lewis2df}
\nocite{stauffer89ic2391}
\nocite{stauffer97}
\nocite{chabrier97}
\nocite{carlsson94}
\nocite{pavlenko96}
\nocite{pavlenko95}
\nocite{bailey02}
\nocite{jones99}
\nocite{allen95}
\nocite{briceno98}
\nocite{allenphd95}
\nocite{montes97}
\nocite{barradom35li01}
\nocite{bessell90}
\nocite{chaboyer95}
\nocite{jeffrieslirich95}
\nocite{hauschildt99}
\nocite{stauffer86}
\nocite{barrado03}
\nocite{oliveira03}
\nocite{smith92}
\nocite{smalley01}
\nocite{randichaperli98}
\nocite{barradoli01}
\nocite{fordpleiades02}
\nocite{meynet97}
\nocite{song02}
\nocite{pinsonneault98}
\nocite{burke00}
\nocite{barrado96}
\nocite{ryan95}
\nocite{zahn94}
\nocite{oppenheimer97}
\nocite{jeffries02}
\nocite{mermilliod81}
\nocite{claria82}
\nocite{bessell91}
\nocite{dejager87}
\nocite{jeffriesn251698}
\nocite{stauffer03}
\nocite{martin94}
\nocite{kirkpatrick91}
\nocite{castelli97}

\bibliographystyle{mn}  
\bibliography{iau_journals,master}  
  
\label{lastpage}  
\end{document}